\documentclass[a4paper,11pt]{article}

\usepackage[svgnames]{xcolor} 

\usepackage[printonlyused,withpage]{acronym} 
\usepackage{amsfonts}
\usepackage{anyfontsize}
\usepackage{bbm}
\usepackage{booktabs} 
\usepackage{braket}
\usepackage{cancel}
\usepackage{changepage}
\usepackage{colortbl}
\usepackage{enumitem}
\usepackage[mathscr]{euscript}
\usepackage{float}
\usepackage[T1]{fontenc} 
\usepackage{gensymb}
\usepackage{Packages/jheppub}
\usepackage{listings} 
\usepackage{lmodern} 
\usepackage{makecell}
\usepackage{mathtools}
\usepackage[framemethod=tikz]{mdframed} 
\usepackage{microtype} 
\usepackage{multirow}
\usepackage{physics}
\usepackage{relsize}
\usepackage{rotating} 
\usepackage{sansmath}
\usepackage{simplewick}
\usepackage{slashed}
\usepackage{stmaryrd}
\usepackage{subfig}
\usepackage{tcolorbox}
\usepackage{theorem}
\usepackage{tikz}
\usepackage{tikz-feynman,contour}
\usepackage[normalem]{ulem}
\usepackage{xspace} 
\usepackage{yfonts}
\usepackage[vcentermath]{youngtab}

\usepackage{siunitx}

\tcbuselibrary{theorems}
\usetikzlibrary{decorations.markings}
\usetikzlibrary{shapes.geometric}

\def\doot{{\boldsymbol{\hspace{0.1em} \cdot\hspace{0.1em}}}}

\makeatletter
\newcommand*{\transpose}{%
  {\mathpalette\@transpose{}}%
}
\newcommand*{\@transpose}[2]{%
  \raisebox{\depth}{$\m@th#1\intercal$}%
}
\makeatother

\newtcbox{\sln}{colback=Gainsboro,
colframe=Gainsboro}

\newcommand{\bt}[1]{{\sansmath{\boldsymbol{#1}}}}

\newcommand{\overbar}[1]{\mkern 2mu\overline{\mkern-4mu#1\mkern-4mu}\mkern 2mu}

\tikzset{snake it/.style={decorate, decoration={snake,amplitude=10mm}}}

\tikzset{/pgf/decoration/.cd,
    number of sines/.initial=10,
    angle step/.initial=20,
}

\newdimen\tmpdimen

\pgfdeclaredecoration{full sines}{initial}
{
    \state{initial}[
        width=+0pt,
        next state=move,
        persistent precomputation={
            \pgfmathparse{\pgfkeysvalueof{/pgf/decoration/angle step}}%
            \let\anglestep=\pgfmathresult%
            \let\currentangle=\pgfmathresult%
            \pgfmathsetlengthmacro{\pointsperanglestep}%
                {(\pgfdecoratedremainingdistance/\pgfkeysvalueof{/pgf/decoration/number of sines})/360*\anglestep}%
        }] {}
    \state{move}[width=+\pointsperanglestep, next state=draw]{
        \pgfpathmoveto{\pgfpointorigin}
    }
    \state{draw}[width=+\pointsperanglestep, switch if less than=1.25*\pointsperanglestep to final, 
        persistent postcomputation={
        \pgfmathparse{mod(\currentangle+\anglestep, 360)}%
        \let\currentangle=\pgfmathresult%
    }]{%
        \pgfmathsin{+\currentangle}%
        \tmpdimen=\pgfdecorationsegmentamplitude%
        \tmpdimen=\pgfmathresult\tmpdimen%
        \divide\tmpdimen by2\relax%
        \pgfpathlineto{\pgfqpoint{0pt}{\tmpdimen}}%
    }
    \state{final}{
        \ifdim\pgfdecoratedremainingdistance>0pt\relax
            \pgfpathlineto{\pgfpointdecoratedpathlast}
        \fi
   }
}

\pgfdeclaredecoration{completely sines}{initial}
{
    \state{initial}[
        width=+0pt,
        next state=upsine,
        persistent precomputation={\pgfmathsetmacro\matchinglength{
            \pgfdecoratedinputsegmentlength / int(\pgfdecoratedinputsegmentlength/\pgfdecorationsegmentlength)}
            \setlength{\pgfdecorationsegmentlength}{\matchinglength pt}
        }] {}
    \state{upsine}[width=\pgfdecorationsegmentlength,next state=downsine]{
        \pgfpathsine{\pgfpoint{0.25\pgfdecorationsegmentlength}{0.5\pgfdecorationsegmentamplitude}}
        \pgfpathcosine{\pgfpoint{0.25\pgfdecorationsegmentlength}{-0.5\pgfdecorationsegmentamplitude}}
    }
    \state{downsine}[width=\pgfdecorationsegmentlength,next state=upsine]{
        \pgfpathsine{\pgfpoint{0.25\pgfdecorationsegmentlength}{-0.5\pgfdecorationsegmentamplitude}}
        \pgfpathcosine{\pgfpoint{0.25\pgfdecorationsegmentlength}{0.5\pgfdecorationsegmentamplitude}}
}
    \state{final}{}
}

\tikzfeynmanset{ with arrow/.style = {
   decoration={
     markings,
     mark=at position 0.5
          with {\arrow[xshift=1.9mm]{Latex[width=2.75mm,length=3mm]}}
     },
   postaction=decorate}
}

\tikzfeynmanset{ with glarrow/.style = {
   decoration={
     markings,
     mark=at position 0.5
          with {\arrow[white,xshift=3mm]{Latex[width=4.125mm,length=4.5mm]}}
     },
   postaction=decorate}
}

\tikzfeynmanset{ with outarrow/.style = {
   decoration={
     markings,
     mark=at position 0.5
          with {\arrow[xshift=3.35mm]{Latex[width=4.58333mm,length=5mm]}}
     },
   postaction=decorate}
}

\tikzfeynmanset{ bigphoton/.style={
    /tikz/draw=none,
    /tikz/decoration={name=none},
    /tikz/postaction={
      /tikz/draw,
      /tikz/decoration={
        completely sines,
        segment length=3mm,
        amplitude=2.5mm,
      },
      /tikz/decorate=true,
    },
  },
}

\tikzfeynmanset{ roundbigphoton/.style={
    /tikz/draw=none,
    /tikz/decoration={name=none},
    /tikz/postaction={
      /tikz/draw,
      /tikz/decoration={
        full sines,
        segment length=3mm,
        amplitude=1.25mm,
      },
      /tikz/decorate=true,
    },
  },
}

\tikzset{ mega thick/.style= {line width = 3.4pt}
}

\makeatletter
\renewcommand{\fnum@figure}{\textsc{\figurename~\thefigure}} 
\makeatother

\lstnewenvironment{pseudoc}{\lstset{frame=lines,language=C,mathescape=true}}{}
\lstnewenvironment{logs}{\lstset{frame=lines,basicstyle=\footnotesize\ttfamily,numbers=none}}{}
\lstnewenvironment{cc}{\lstset{frame=lines,language=C}}{}
\lstnewenvironment{c64}{\lstset{backgroundcolor=\color{c64},basewidth=0.65em,basicstyle=\commodoreface\color{c64light},numbers=none,framerule=10pt,rulecolor=\color{c64light},frame=tb,framexbottommargin=30pt}}{}
\lstnewenvironment{html}{\lstset{frame=lines,language=html,numbers=none}}{}
\lstnewenvironment{pseudo}{\lstset{frame=lines,mathescape=true,morekeywords={learn_string_domain, save_model}}}{}
\lstnewenvironment{pseudoctiny}{\lstset{language=C,mathescape=true,basicstyle=\tiny\sffamily}}{}
\lstnewenvironment{cctiny}{\lstset{language=C,basicstyle=\tiny\sffamily}}{}
\lstnewenvironment{pseudotiny}{\lstset{mathescape=true,basicstyle=\tiny\sffamily}}{}

\title{LEX-EFT: The Light Exotics Effective Field Theory}

\author{Linda M. Carpenter,}
\author{Taylor Murphy,}
\author{and Matthew J. Smylie}
\affiliation{Department of Physics, The Ohio State University\\ 191 W. Woodruff Ave., Columbus, OH 43210, U.S.A.}

\emailAdd{lmc@physics.osu.edu}
\emailAdd{murphy.1573@osu.edu}
\emailAdd{smylie.8@osu.edu}

\date{\today}

\abstract{\begin{abstract}

We propose the creation of a Light Exotics Effective Field Theory (LEX-EFT) catalog. LEX-EFT is a generic framework to capture all interactions between the Standard Model (SM) and all (or at least a large class of) theoretically allowed exotic states beyond the Standard Model (bSM), indexed by their SM and bSM charges. These states are light enough to be on or nearly on shell in some collider processes. This framework, which subsumes beyond the Standard Model paradigms as generally as possible, is meant to extend recent successful implementations of bSM EFTs and complement \emph{e.g.} the Standard Model Effective Field Theory (SMEFT), which can capture the off-shell effects of exotic fields. In this work, we review a general method for the construction of a complete list of gauge-invariant operators involving SM interactions with light exotics via iterative tensor product decomposition, up to the desired order in mass dimension. Each operator is characterized by specific Clebsch-Gordan coefficients determined by the charge flow; we show how this charge flow affects the range of EFT validity and cross sections associated with an effective operator. We create an example catalog of exotic scalars coupling to SM gauge boson pairs, and we highlight some operators with exotic weak $\mathrm{SU}(2)_{\text{L}}$ charges that can produce spectacular LHC phenomenology. We further demonstrate the utility of the LEX-EFT approach with several examples of effects on kinematic distributions and cross sections that would not be captured by EFTs agnostic to the exotic degrees of freedom and may evade the main inclusive collider searches tailored to the existing preferred set of standard bSM theories.

\end{abstract}}

\begin{document}

\maketitle

\section{Introduction}

 As the LHC era continues, it is important to leave no stone unturned in the search for new phenomena beyond the Standard Model (bSM). The space of all possible phenomenological signatures is vast, and new methods are needed to ensure that a maximal region of this space is being explored. In recent years, new ground has been opened up by the introduction of general approaches. Computational techniques like anomaly detection, for example, remain model agnostic \cite{Kasieczka:2021xcg}. Along different lines, the Standard Model Effective Field Theory (SMEFT) attempts to capture new physics by enumerating a complete set of general operators that govern Standard Model interactions, with all new states assumed heavy (off shell) and integrated out of the theory \cite{Grzadkowski_2010}. 
 
 The use of effective field theories (EFTs) has proven to be a phenomenological strategy of great utility in the LHC era, as it promises to cast the widest possible net in the pursuit of new physics. EFTs offer a simple formalism for cataloging the interactions of SM and bSM states. Much work has been done within the SMEFT framework, mentioned just above, which at the time of writing has a complete catalog of SM operators up to dimension eight and includes Higgs interactions. This approach has led to a plethora of collider analyses intended to measure and constrain the SMEFT Wilson coefficients in order to search for new physics. Discoveries of or constraints on bSM states based on SMEFT analyses requires matching between the effective theory and a full bSM model. Recent applications of EFTs beyond the Standard Model include a diverse and vibrant dark matter (DM) program \cite{Goodman:2010ku,Goodman:2010yf,Alanne:2017oqj,Goertz:2019wtt,Alanne:2020xcb} and interactions of axion-like particles \cite{Brivio:2017ije}. The DM EFT catalog, for example, has led to the exploration of many new possible dark matter discovery channels at the LHC \cite{Belotsky:1998pv}. 

Motivated by these successes, we seek to expand the theoretical coverage of the phenomenological landscape by introducing a \emph{Light Exotics Effective Field Theory} (LEX-EFT), a systematized general approach to interactions of exotic states with the SM. Light exotic fields are those light enough not to be totally integrated out of the theory. Since LEX particles can appear on shell in at least some considered phenomenological processes, LEX-EFT is complementary to the consideration of the off-shell processes carried out in \emph{e.g.} the SMEFT. 

In this work, to be precise, we propose the creation of a comprehensive effective operator catalog of interactions between exotic states and the Standard Model, indexed by the specific gauge representations of the exotic fields. Ideally, LEX-EFT keeps the concision and generality that makes the EFT approach to phenomenology advantageous. In particular, we view the advantages of LEX-EFT as follows:
\begin{itemize}
\item LEX-EFT offers a complete list of all possible interactions between light exotics and the Standard Model up to the desired order in effective cut-off (mass dimension). It is thus a guide for bSM precision and collider searches, it allows for the analysis of new event topologies, and it offers a comprehensive map of event kinematics without the burden of specifying UV-complete models.

\item A complete LEX-EFT catalog would subsume other classes of exotic bSM models including supersymmetry, exotic Higgs models, and dark matter EFTs. Such a complete catalog may illuminate new interactions in these theories and thus new phenomenological channels for study.

\item The LEX-EFT catalog would also bring to theoretical consideration bSM states that have not received model-building attention. It would thus cast a wider net over all of theory space in a systematic manner, accomplishing a goal that in the past few years has crystallized and started to receive attention from the theory community \cite{Banerjee:2020bym,Banerjee:2020jun,Li:2022tec}. As we imagine the LEX-EFT approach would be closely followed up by a simplified model building approach, this would spark new theoretical innovation.
\end{itemize}

The LEX-EFT approach removes some of the model agnosticism of other general approaches to phenomenology, yet it allows the capture of many phenomenological features of collider processes that would not be possible otherwise. We highlight the following distinctive features in this work:
\begin{itemize}
\item \emph{Kinematics and collider cross sections}: Using unique LEX operators allows one to keep track of process kinematics, which are vital in constructing collider searches for new physics. It also allows for the accurate computation of collider production cross sections, scaled by the relevant effective operator coefficients, up to the validity limits of the EFT. This allows full consideration of all processes involving production and decay of exotic states in collider searches.

\item \emph{Charge flow and validity of parameter space}: Constructing effective operators that are singlets under all gauge groups requires specification of the Clebsch-Gordan coefficients in operators linking light exotic fields to the SM. For any given set of fields, there may be multiple ways to perform charge contractions. Each of these contractions then corresponds to a unique operator, which gives a picture of the charge flow of the process involved. There may be naturally large coefficients associated with some operators, which drastically affects predictions of production cross sections in the theory. Moreover, we find that the range of validity of an effective operator may vary widely based on choice of charge contraction, even if the fields involved in the operators are the same.
\end{itemize}
Even though the LEX-EFT approach focuses on the collider production of on-shell new states, the proposed operator catalog does have implications for loop-level processes, which should be explored in future work. In particular:
\begin{itemize}
\item \emph{Operator correlations}: A theory containing a specific LEX state has operators that may have correlations based on gauge invariance or other theoretical considerations. This approach works even with  LEX states that are totally off-shell. The operator catalog for off-shell states leads to a specific list of correlated SMEFT operators that could be measured once the bSM states are integrated out.

\item \emph{Precision measurements at loop level}: Specifying the light exotic state appearing in a theory facilitates the computation of precision quantities such electroweak oblique parameters, lepton anomalous magnetic dipole moments, $b \to s \gamma$, etc., which may not be obvious from other operator catalogs.
\end{itemize}

This paper is organized as follows. In \hyperref[s2]{Section 2} we review the iterative construction of Lorentz- and gauge-invariant operators including light exotic fields. We moreover introduce an example catalog of operators featuring scalars in higher-dimensional representations of $\mathrm{SU}(2)_{\text{L}}$. In \hyperref[s3]{Section 3} we explore the idea of exotic charge contraction and demonstrate how the quantum numbers of light exotics can affect both LHC cross sections for (b)SM processes and the valid experimentally accessible EFT parameter space. In \hyperref[s4]{Section 4}, we provide two phenomenological examples of light-exotics models that produce identical final states at the LHC but exhibit totally distinct kinematics that cannot be captured in any EFT that excludes bSM degrees of freedom. \hyperref[s5]{Section 5} summarizes this work and suggests avenues of future research within the LEX-EFT framework.
\section{An iterative tensor product method to construct new singlet operators} 
\label{s2}


The LEX-EFT framework is underpinned by a straightforward group-theoretic procedure for obtaining a complete operator list of novel gauge-singlet operators up to a specified dimension. We therefore begin this work by describing the procedure in general and providing some reasonably self-contained examples. We start with a new LEX state denoted by $\Phi_i$ that lies in a specified representation of SM and bSM gauge/global groups. The goal is to create a complete catalog of singlet operators that couple these LEX fields and SM fields $\psi_i$ up to a given order in an EFT cutoff $\Lambda$. More precisely, each effective operator has the form
\begin{equation}
    \frac{1}{\Lambda^d}\,\lambda \,(\Phi_1 \Phi_2 \dots)(\psi_1 \psi_2 \dots)
\end{equation}
plus first derivatives of such fields\footnote{We forbid higher-derivative operators in order to avoid the unbounded Hamiltonians predicted by Ostrogradsky's theorem \cite{Motohashi:2020psc}.}, and we must find a complete list of all charge singlets composed of SM and LEX fields up to the desired mass dimension $4+d$. Further, the coupling coefficients $\lambda$ contain the group-theoretic information about how to complete the charge contraction of the fields in the operator. As we discuss in \hyperref[s3]{Section 3}, there may be more than one way to contract the charges, and in general there are many distinct charge contractions of multi-field operators. Thus the operator coefficients are different, and we must consider these separate operators.

In order to make sure our list of charge singlets is complete, we follow an iterative procedure that exploits the known group theory tensor products of irreducible representations of semisimple Lie groups. We consider the fields $\Phi_i$ and $\psi_j$ to be in some irreducible representation(s) of these groups, with an $r$-dimensional representation denoted in general by $\textbf{r}$.\footnote{This simple notation can cause trouble when there are multiple distinct irreducible representations with equal dimension, as for the $\boldsymbol{15}$ and $\boldsymbol{15'}$ of $\mathrm{SU}(3)$. This specific problem does not impact any results in this work, but there is standard notation to distinguish between such representations \cite{Georgi:1982jb}.}  We can then consider the representation of the direct product of pairs of fields,
\begin{equation}
    \textbf{r}_1 \otimes \textbf{r}_2 = \textbf{q}_1 \oplus \textbf{q}_2 \oplus \dots.
\end{equation}
For any given group there exists a list of such tensor products. We now discuss the construction of invariants from this list of bilinear tensor products.

\emph{Observation}. If there exist invariant combinations of $n+1$ and $m+1$ fields transforming in the direct product representations $\textbf{r}_1 \otimes \cdots \otimes \textbf{r}_n \otimes \textbf{p}$ and $\textbf{q}_1 \otimes \cdots \otimes \textbf{q}{}_{m} \otimes \textbf{p}$ of a group, then there exists an invariant combination of $n+m$ fields in the reducible representation $\textbf{r}_1 \otimes \cdots \otimes \textbf{r}_n \otimes \boldsymbol{\bar{\textbf{q}}}_1 \otimes \cdots \otimes \boldsymbol{\bar{\textbf{q}}}{}_{m}$ \cite{Carpenter:2021rkl}.

\emph{Example}. Suppose that two distinct tensor products contain the same irreducible representation \textbf{e}; that is, that
\begin{equation}
    \mathbf{a}\otimes \mathbf{b} \supset \mathbf{e}  \ \ \text{and}\ \ \mathbf{c}\otimes \mathbf{d} \supset \mathbf{e}.
\end{equation}
In this case we immediately infer the existence of the two trilinear invariants
 \begin{equation}
\mathbf{{a}\otimes b \otimes \boldsymbol{\bar{\textbf{e}}}} \ \ \text{and}\ \ \ \textbf{c} \otimes \textbf{d} \otimes \boldsymbol{\bar{\textbf{e}}},
 \end{equation}
and we can also create a new iterated invariant by exploiting the fact that $\textbf{e} \otimes \boldsymbol{\bar{\textbf{e}}}$ contains a singlet: 
\begin{equation}
    \mathbf{a} \otimes \mathbf{b} \otimes \boldsymbol{\bar{\textbf{c}}} \otimes \boldsymbol{\bar{\textbf{d}}} \supset \boldsymbol{1}.
\end{equation}
This singlet contains the direct product of four irreducible representations, which can be mapped back to an operator containing SM and/or LEX fields. We note that the ``intermediate representation'' $\mathbf{e}$ need not be in a representation corresponding to the fields in the theory. It may, however, be useful in determining the flow of charge.

This process can be iterated with further nesting of bilinear tensor products in order to create singlets containing five representations, then six, etc. By continuing in this manner and mapping onto (b)SM fields, we can create gauge-invariant operators with more states. The iterative insertion of tensor products may be systematized to create complete lists of invariants which contain a specified number $N$ of LEX/SM fields in irreducible representations. We refer to such singlets as ``$N$-field invariants''. The lists of $N$-field invariants will be complete as long as all possible intermediate states (representations) are accounted for. The complete list of invariants can then become a list of effective operators with predetermined field content up to the desired order in an effective field theory expansion (\emph{e.g.} in the cutoff $\Lambda^{-1}$). 
 
For example, suppose we wanted to create a complete list of invariants containing four fields. We would begin by noting the representations of the SM or LEX fields, thus mapping $\{\Phi_i, \psi_i\} \rightarrow \mathbf{r}_i$. We would then determine all possible bilinear tensor products that involve these representations. From here we can create a list of three-state invariants. We then follow the iterative expansion process above, inserting all possible bilinears in intermediate states, to create a \emph{complete} list of singlet products containing four terms. These can then be mapped back into the SM or LEX states to create a \emph{complete} list of operators. The operators obtained by mapping back onto the states $\{\Phi_i,\psi_i\}$ will be proportional to the Clebsch-Gordan coefficients that contract the specific charge indices of the four-state invariants.

Though the list of $N$-field invariants is complete, the resulting terms contain fields of various mass dimension (gauge field-strength tensors, scalars, spin 1/2 fermions, etc) and therefore may map to operators of various effective dimension. Nevertheless, a similar iterative process may be employed to create lists with five fields, six fields, etc. Eventually all operators up to the desired dimension in EFT will be found. We give a brief argument below concerning the completeness of this process. 

\emph{Regarding completeness}. There exist in the theory a finite number, $M$, of SM/LEX fields transforming in irreducible representations $\mathbf{r}_i$ with $i\in\{1,\dots, M\}$. An invariant of interest contains a specified number $N$ of these fields. Since we are concerned with invariants involving $N$ fields, we must contract all indices. The intermediate sub-product of a LEX state in representation $\textbf{r}_{\text{LEX}}$ with any other states of the theory will be in some intermediate representation $\mathbf{r'}$. The sub-product of the remaining states must contracted in the conjugate representation $\boldsymbol{\bar{\textbf{r}}'}$, so that singlets take the form
\begin{equation}
    \underbrace{\left[\mathbf{r}_{\text{LEX}}\otimes \mathbf{r}_{i} \otimes \mathbf{r}_{j} \otimes \dots \otimes \mathbf{r}_{k} \right]_\mathbf{r'}}_{N-n} \otimes \underbrace{ \left[\mathbf{r}_{l} \otimes \dots \otimes \mathbf{r}_{m} \right]_{\boldsymbol{\bar{\textbf{r}}'}}}_{n}\ \ \ \text{with}\ \ \ 1\leq i,j,\dots,m\leq M-1.
\end{equation}
There will be a maximal representation size to any sub-product of states within the $N$-field invariant, hence a maximum representation size of any intermediate representation $\mathbf{r'}$.

We now argue from induction. To build three-field invariants involving a LEX field, we need only consider the $m$ possible bilinear tensor products of the LEX state with other representations allowed in the theory, $[\mathbf{r}_{\text{LEX}}\otimes \mathbf{r}_i]_{\boldsymbol{\textbf{r}}'_j}$, to obtain the finite list of irreducible representations $\mathbf{r'}$ in the direct product. If any single field in the theory is in the conjugate representation $\boldsymbol{\bar{\textbf{r}}}'_j$, then we can directly contract indices to form an invariant:


\begin{equation}
   [\mathbf{r}_{\text{LEX}}\otimes \mathbf{r}_i]_{\textbf{r}'_j} \otimes \boldsymbol{\bar{\textbf{r}}}'_j
\end{equation}
With a list in hand of all $m$ possible bilinear products $\mathbf{r}_{\text{LEX}}\otimes \mathbf{r}_i$ in representations $\boldsymbol{\textbf{r}}'_j$, we can proceed to construct the four-field invariants. We find the direct products of the allowed representations $\mathbf{r}_k \otimes \mathbf{r}_l$ that are in a given conjugate representation $\boldsymbol{\bar{\textbf{r}}}'_j$ and contract these fields according to
\begin{equation}
   [ \mathbf{r}_{\text{LEX}}\otimes \mathbf{r}_i]_{\boldsymbol{\textbf{r}}'_j} \otimes [\mathbf{r}_{k} \otimes { \mathbf{r}_l}]_{\boldsymbol{\bar{\textbf{r}}}'_j}
\end{equation}
to obtain singlets.
To proceed to five fields, we now consider all possible trilinear products of the form $\mathbf{r}_{\text{LEX}}\otimes \mathbf{r}_i \otimes \mathbf{r}_j$.
We note we have already found by exhaustion the representations of bilinear products of the first two fields in the previous step. In that step, the bilinears were in representations $\mathbf{r}'_j$ such that $\mathbf{r}_{\text{LEX}}\otimes \mathbf{r}_i\supset \mathbf{r}'_j$. We can thus iterate the bilinear tensor products $\mathbf{r}'_j\otimes \mathbf{r}_j\supset \mathbf{r}'_k$ to find the representations $\mathbf{r}'_k$ of all trilinear products. We then find the remaining bilinear representations $\mathbf{r}_k \otimes \mathbf{r}_l$ that are in the conjugate representation $\boldsymbol{\bar{\textbf{r}}}'_k$ and contract \emph{these} fields to form the five-field invariant. This process can be repeated indefinitely and will ultimately produce all possible terms --- we only need to know the list of bilinear tensor products that involve relevant SM/LEX fields and the intermediate representations $\mathbf{r}'_j$, $\mathbf{r}'_k$, and so on. We note that this method can be applied not only to constructing gauge singlets but also straightforwardly to representations of the Lorentz group, since fields (and their first derivatives) in irreducible representations of the Lorentz group can be characterized by $\mathrm{SU}(2) \times \mathrm{SU}(2)$ quantum numbers.

We can choose different ways to build the effective operator catalog. One way is, given a specific field content, to build all operators containing a certain number of fields. Another is to build all operators up to a certain dimension in the EFT cutoff $\Lambda$. Yet another way is to build all operators that interact through a certain portal. There are some good application of this in studies of dark matter where a DM candidate may interact, for example, through a quark portal, so that all possible gauge-singlet DM-quark operators should be specified.

Note that in order for a model to be a true EFT, it must contain every possible operator up to a specified mass dimension. Generating complete bases of independent operators for EFTs has been the subject of much research \cite{Grzadkowski_2010,Lehman:2015via,Henning:2015daa,Henning:2017fpj}, involving both traditional group theory constructions and the Hilbert series method, and there exist computational tools \cite{Gripaios:2018zrz,Criado:2019ugp,Fonseca:2019yya} for listing invariant products of fields. There are some complications when (covariant) derivatives are present; namely, some seemingly different operators may be related via integration by parts, and some operators may vanish when equations of motion for the fields are enforced. In the following sections, we consider extending the SM by adding scalars in many different representations of the SM gauge group. A full list of invariant operators for each model is beyond the scope of this work. Instead, we take a signal-based approach, and we choose to outline only those operators that can result in diboson resonances. Any of these models could be promoted to a complete EFT in order to take full advantage of the formalism, and we leave such efforts to future work.

\subsection{Example: A catalog of exotic scalars in the diboson portal}

The rest of this section is devoted to providing an example LEX-EFT catalog of operators that produce novel phenomenology. This operator list extends to mass dimension seven. We begin with a simple phenomenological idea: we wish to catalog all couplings between a CP-even spin-0 field and pairs of SM gauge bosons. These new LEX scalars $\phi$ can carry various SM quantum numbers but are restricted to be singlets in any bSM gauge groups. Our example catalog serves two demonstrative purposes. First, it gives further practice in using the tensor product technique to produce novel singlet operators; second, it demonstrates that a simple idea --- the diboson portal coupling to a single scalar --- can give rise to disparate and novel event topologies.  

The complete list of operators is found in Tables \hyperref[operator-table]{1}--\hyperref[operator-table-4]{4}, which are organized by the mass dimension of the operators. Again, we list operators up to dimension seven, including insertions of Higgs fields. For any operator that contains a Higgs insertion, the Higgs field may be set to its vacuum expectation value, lowering the effective mass dimension of the operator at a cost of a $v/\Lambda$ suppression. We have written CP-preserving terms only. In the left columns we list the LEX scalar field quantum numbers under the SM gauge group $\mathrm{SU}(3)_{\text{c}} \times \mathrm{SU}(2)_{\text{L}} \times \mathrm{U}(1)_Y$. The right columns contain the effective operators falling under each previous category. 

Let us first discuss LEX states with $\mathrm{SU}(3)_{\text{c}}$ quantum numbers. In order to maintain gauge invariance, operators that contain a single gluon field-strength tensor $G_{\mu\nu}$ must necessarily contain a LEX field $\phi$ that is in the adjoint representation ($\boldsymbol{8}$) of $\mathrm{SU}(3)_{\text{c}}$. So-called color octets appear in many bSM scenarios, such as SUSY and minimal flavor violation (MFV), and exhibit interesting and varied phenomenology \cite{Manohar:2006ga,Plehn:2008ae,Carpenter:2020evo,Carpenter:2020hyz}. Diboson couplings of color octets, in particular, do appear in the literature \cite{Carpenter:2015gua, Carpenter:2021gpl,Hayreter:2018ybt} but are underdiscussed, mainly focusing on the digluon coupling and resultant dijet resonances. A color-octet LEX state might be a singlet under $\mathrm{SU}(2)_{\text{L}}$ or have nontrivial weak quantum numbers. Color octets with $\mathrm{SU}(2)_{\text{L}}$ quantum numbers are quite interesting but have received far less phenomenological attention than weak-singlet color octets. For instance, a weak-doublet color octet with SM quantum numbers $(\boldsymbol{8},\boldsymbol{2},\tfrac{1}{2})$ was proposed in the Manohar-Wise model \cite{Manohar:2006ga}, and produces some interesting collider signatures \cite{Carpenter:2011yj,Hayreter:2018ybt}. Yet this model is still understudied, as the masses of these fields are still quite unconstrained by collider searches. LEX fields with these quantum numbers may couple to $W^{\mu\nu}G_{\mu\nu}$ with the addition of one Higgs insertion to create a SU(2) singlet; namely,
\begin{align}
    \mathcal{L} \supset \frac{1}{\Lambda^2}\, H^{\dagger i}(\sigma^a)_i^{\ j}\phi^A_j W^{a\mu\nu}G^A_{\mu\nu},
    \end{align}
with $\sigma^a$ (at least proportional to) the generators of the fundamental representation of $\mathrm{SU}(2)_{\text{L}}$ such that $a$ is a weak adjoint index and $i,j$ are fundamental indices (see \hyperref[operator-table]{Table 1} and following for index conventions). Similarly, the \emph{biadjoint} field with SM quantum numbers $(\boldsymbol{8},\boldsymbol{3},0)$ has only been studied, to our knowledge, in the context of electroweak oblique corrections \cite{Carpenter:2022oyg}. Within the LEX-EFT framework, such a field may couple to $W^{\mu\nu}G_{\mu\nu}$ through the dimension-five operator
\begin{align}
    \mathcal{L} \supset \frac{1}{\Lambda}\,\phi^{Aa}\,W^{a\mu\nu}G^A_{\mu\nu}.
\end{align}
  We do note that at dimension seven, even the standard weak-singlet color-octet $(\boldsymbol{8},\boldsymbol{1},0)$ scalar may couple to $W^{\mu\nu}G_{\mu\nu}$ through the operator
  \begin{align}
       \mathcal{L} \supset \frac{1}{\Lambda^3}\,(H^\dagger \sigma^a H)\,\phi^A\, W^{a\mu\nu}G^A_{\mu\nu}.
      \end{align}
    It is also possible for color octets in the quadruplet ($\boldsymbol{4}$) and quintuplet ($\boldsymbol{5}$) representations of $\mathrm{SU}(2)_{\text{L}}$ to couple to the diboson pairs $W^{a\mu\nu}G^A_{\mu\nu}$ through operators with additional Higgs insertions. These operators are of particular interest because they contain multiply-electrically-charged states.

    \renewcommand{\arraystretch}{1}
\begin{table*}[]
    \centering
Mass dimension 5 [$\times 1/\Lambda$]\\[1ex]
    \begin{tabular}{c||c}  \toprule\hline
        \rule{0pt}{3.5ex}\ \ $\mathrm{SU}(3)_{\text{c}}\times \mathrm{SU}(2)_{\text{L}}\times \mathrm{U}(1)_Y$\ \ \ & Operators \\[1ex]
        \hline
        \hline
        \rule{0pt}{3.5ex}\multirow{4}{*}[-3.25ex]{$(\boldsymbol{1},\boldsymbol{1},0)$} & $\phi\, B^{\mu \nu} B_{ \mu \nu}$ \\[1ex]\cline{2-2}
        \rule{0pt}{3.5ex} & $\phi\,W^{\mu \nu,a} W^a_{ \mu \nu}$\\[1ex]\cline{2-2}
        \rule{0pt}{3.5ex} & $\phi\, G^{\mu \nu} G_{ \mu \nu}$\\[1ex]\cline{2-2}
        \rule{0pt}{3.5ex} & $ \phi\, (D^{ \mu}H)^{\dagger i}(D_{\mu}H)_i $ \\[1ex]
        \hline
        \rule{0pt}{3.5ex} \multirow{2}{*}[-1.4ex]{$(\boldsymbol{8},\boldsymbol{1},0)$} & $d^{ABC}\,\phi_A\, G_B^{\mu \nu} G_{\mu \nu C}$ \\[1ex] \cline{2-2}
        \rule{0pt}{3.5ex} & $\phi_{A}\, G^{\mu \nu A} B_{ \mu \nu}$ \\[1ex]
        \hline
        \rule{0pt}{3.5ex} \multirow{2}{*}[-1.4ex]{$(\boldsymbol{1},\boldsymbol{3},0)$} & $\phi^{a}\,W^{\mu\nu,a}B_{\mu\nu}$\\[1ex]\cline{2-2}
        \rule{0pt}{3.5ex} & $ \phi_{ij}\, (D^{\mu}H)^{\dagger i} (D_{ \mu}H)^{ j}$ \\[1ex]
        \hline
        \rule{0pt}{3.5ex} $(\boldsymbol{8},\boldsymbol{3},0)$ & $\phi^{a}\,W^{\mu\nu,a}G_{\mu\nu}$\\[1ex] \hline
        \rule{0pt}{3.5ex} $(\boldsymbol{1},\boldsymbol{5},0)$ & $\phi_{ijkl}\,W^{ij\mu\nu}W^{kl}_{\mu\nu}$ \\[1ex]
        \hline
        \rule{0pt}{3.5ex}  $(\boldsymbol{10},\boldsymbol{1},0)$ & $ \varepsilon^{KLM}\phi_{IJK}\, G_L^{I\mu\nu}G^J_{M\mu\nu}$ \\[1ex]
        \hline
        \rule{0pt}{3.5ex}  $(\boldsymbol{27},\boldsymbol{1},0)$ & $ \phi_{IJ}^{KL}\, G_K^{I\mu\nu}G^J_{L\mu\nu}$ \\[1ex]
        \hline
        \bottomrule
    \end{tabular}
    \caption{Dimension-five operators that couple boson pairs to bSM fields $\phi$ with specified SM quantum numbers. Here $\mathrm{SU}(2)_{\text{L}}$ indices ($i,j,\dots$ fundamental and $a,b,\dots$ adjoint) are lowercase and $\mathrm{SU}(3)_{\text{c}}$ indices are capital letters.}
    \label{operator-table}
\end{table*}

LEX fields that couple to a pair of gluon field strengths $G_{\mu\nu}G^{\mu\nu}$ may be in various representations. With the decomposition
\begin{align}
    \boldsymbol{8} \otimes \boldsymbol{8} &= \boldsymbol{1}_{\text{s}} \oplus \boldsymbol{8}_{\text{s}} \oplus \boldsymbol{8}_{\text{a}} \oplus \boldsymbol{10}_{\text{a}} \oplus \boldsymbol{\overbar{10}}_{\, \text{a}} \oplus \boldsymbol{27}_{\text{s}},
\end{align}
we see the LEX state may be in a singlet, adjoint, decuplet ($\boldsymbol{10}$), or $\boldsymbol{27}$ of $\mathrm{SU}(3)_{\text{c}}$. The LEX states in these operators may appear at dimension five with $\mathrm{SU}(2)_{\text{L}} \times \mathrm{U}(1)_Y$ quantum numbers $(\boldsymbol{1},1)$; they may appear at dimension six with $\mathrm{SU}(2)_{\text{L}} \times \mathrm{U}(1)_Y$ quantum numbers $(\boldsymbol{2},\tfrac{1}{2})$ via the insertion of one Higgs state, and they may appear at dimension seven with two Higgs insertions in the weak triplet $(\boldsymbol{1},1)$ or singlet $(\boldsymbol{1},1)$ representations. We note that a field in the $\boldsymbol{10}$ representation can be written as a symmetric tensor with three fundamental indices, and one in the $\boldsymbol{27}$ can be written as a symmetric tensor with two fundamental and two anti-fundamental indices. We make use of this notation in Tables \hyperref[operator-table]{1} and \hyperref[operator-table-4]{4}.

\subsection{Higher-dimensional representations of $\text{SU}(2)_{\text{L}}$}

To elaborate upon this example, we now discuss the construction of operators involving various representations of the weak $SU(2)_L$ gauge group. It is well known that the representations of $\mathrm{SU}(2)$ may be mapped onto simple spin algebra from quantum mechanics, where the $n$-dimensional representation maps onto objects of spin $J$ with $n=2J+1$. Thus, for example, a field in the five-dimensional representation maps to a spin-2 object with five possible spins: $J \in \{-2,-1,0,1,2\}$. We may then infer the tensor product relations among operators containing fields charged under $\mathrm{SU}(2)$. Recall that the tensor products of objects with spins $J$ and $L$ with $J \geq L$ follow
\begin{align}
J \otimes L = J+L \oplus J+L-1 \oplus \dots \oplus J-L.
\end{align}
As an example, consider the tensor product of two three-dimensional representations of $\mathrm{SU}(2)$, $\boldsymbol{3}\otimes \boldsymbol{3}$. The triplets of SU(2) map to $J=L=1$, so the possible spin-product states are $J \in \{0,1,2\}$, corresponding to the one-, three-, and five-dimensional representations of SU(2). We therefore arrive at the tensor product relation
\begin{align}
    \boldsymbol{3} \otimes \boldsymbol{3} = \boldsymbol{1} \oplus \boldsymbol{3} \oplus \boldsymbol{5}
\end{align}
in SU(2). From here, we can use the iterative tensor product method to construct the singlet operators that couple LEX states in higher-dimensional representations of SU(2) to pairs of gauge bosons.  

\begin{table*}[]
    \centering
    Mass dimension 6 [$\times 1/\Lambda^2$]\\[1ex]
    \begin{tabular}{c||c} 
    \toprule\hline
        \rule{0pt}{3.5ex} \ \ $\mathrm{SU}(3)_{\text{c}}\times \mathrm{SU}(2)_{\text{L}}\times \mathrm{U}(1)_Y$\ \ \ & Operators \\[1ex]
        \hline
        \hline
        \rule{0pt}{3.5ex}\multirow{5}{*}[-4.5ex]{$(\boldsymbol{1},\boldsymbol{2},-\frac{1}{2})$} & $H^{ i}(\sigma^a)_i^{\ j}\phi_j\, W^{\mu\nu,a}B_{\mu\nu}$ \\[1ex]\cline{2-2}
        \rule{0pt}{3.5ex} & $[H^{ i}  \phi_i]\, B^{\mu \nu} B_{ \mu \nu}$ \\[1ex]\cline{2-2}        
        \rule{0pt}{3.5ex} & $[H^{ i}  \phi_i]\, W^{\mu \nu,a} W^a_{ \mu \nu}$ \\[1ex]\cline{2-2}        
        \rule{0pt}{3.5ex} & $[H^{ i}  \phi_i]\, G^{\mu \nu} G_{ \mu \nu}$ \\[1ex] \cline{2-2}
        \rule{0pt}{3.5ex} & $ [H^{ i}  \phi_i]\,D^{\mu}H_j D_{ \mu}H^{\dagger j}$ \\[1ex]
        \hline
        \rule{0pt}{3.5ex}\multirow{3}{*}[-2.0ex]{$(\boldsymbol{8},\boldsymbol{2},-\frac{1}{2})$} & $d^{ABC}\, [H^{ i} \phi_{Ai}]\, G_{B}^{\mu \nu}G_{C\mu\nu}$ \\[1ex]\cline{2-2}
        \rule{0pt}{3.5ex}  & $[H^{ i} \phi_{Ai}]\, G_{A}^{\mu \nu}B_{\mu \nu}$ \\[1ex]\cline{2-2}
        \rule{0pt}{3.5ex}  & $H^{ i}(\sigma^a)_i^{\ j}\phi_j\, W^{\mu\nu,a}G_{\mu\nu} $ \\[1ex] \cline{2-2}
        \hline
        \rule{0pt}{3.5ex}\multirow{3}{*}[-2.0ex]{$(\boldsymbol{1},\boldsymbol{4},-\frac{1}{2})$} & $\phi_{ijk}\,H^k\, W^{ij\mu\nu}B_{\mu\nu}$ \\[1ex]  \cline{2-2}
        \rule{0pt}{3.5ex} & $\phi_{ijk}\,H_l \,W^{ij\mu\nu}W^{kl}_{\mu\nu}$ \\[1ex] \cline{2-2}
        \rule{0pt}{3.5ex} & $[H^{ i}  \phi_{ijk}]\,(D^{\mu}H)^{j} (D_{ \mu}H)^{\dagger k}$ \\[1ex] \hline
        
        \rule{0pt}{3.5ex}$(\boldsymbol{8},\boldsymbol{4},-\frac{1}{2})$ & $\phi_{ijk}\,H^k\, W^{ij\mu\nu}G_{\mu\nu}$ \\[1ex] \hline
        \rule{0pt}{3.5ex}$(\boldsymbol{1},\boldsymbol{6},-\frac{1}{2})$ & $\phi_{ijklm}\,H^m \,W^{ij\mu\nu}W^{kl}_{\mu\nu}$ \\[1ex]
    \hline
    \bottomrule
    \end{tabular}
\caption{Dimension-six exotic operators that couple boson pairs to BSM fields $\phi$ with specified SM quantum numbers. Indices are as shown in \hyperref[operator-table]{Table 1}.}
    \label{operator-table-2}
\end{table*}
\renewcommand{\arraystretch}{1.0}

\begin{table*}[]
    \centering
Mass dimension 7 [$\times 1/\Lambda^3$]\\[1ex]
\scalebox{0.89}{    \begin{tabular}{c||c}
    \toprule\hline
        \rule{0pt}{3.5ex} \ \ $\mathrm{SU}(3)_{\text{c}}\times \mathrm{SU}(2)_{\text{L}}\times \mathrm{U}(1)_Y$\ \ \ & Operators \\[1ex]
        \hline
        \hline
        \rule{0pt}{3.4ex}\multirow{9}{*}[-9ex]{$(\boldsymbol{1},\boldsymbol{1},0)$} & $|H|^2 \phi\, B^{\mu \nu} B_{ \mu \nu}$ \\[1ex] \cline{2-2}
        \rule{0pt}{3.5ex} & $|H|^2 \phi\, W^{\mu \nu,a} W^a_{ \mu \nu}$\\[1ex] \cline{2-2}
        \rule{0pt}{3.5ex} & $|H|^2 \phi\, G^{\mu \nu} G_{ \mu \nu}$\\[1ex] \cline{2-2}
        \rule{0pt}{3.5ex} & $(H^\dagger \sigma^a H)\,\phi\, W^{\mu\nu,a}B_{\mu\nu}$ \\[1ex] \cline{2-2}
        \rule{0pt}{3.5ex} & $D^{\mu}\phi D^{\nu}H^\dagger_iH_{j}\,W^{ij}_{\mu\nu}$ \\[1ex] \cline{2-2}
        \rule{0pt}{3.5ex} & $D^{\mu}\phi D^{\nu}H_iH^{\dagger i}\,B_{\mu\nu}$ \\[1ex] \cline{2-2}
        \rule{0pt}{3.5ex} & $ \phi\, |H|^2 |D_\mu H|^2 $ \\[1ex] \cline{2-2}
        \rule{0pt}{3.5ex} & $ \phi\, |H^\dagger D_\mu H|^2 $ \\[1ex]\cline{2-2}
        \rule{0pt}{3.5ex} & $[H^{\dagger}H  \phi]\,(D^{\mu}H)( D_{ \mu}H^\dagger) $ \\[1ex]
        \hline
        \rule{0pt}{3.5ex}\multirow{12}{*}[-12ex]{$(\boldsymbol{1},\boldsymbol{3},0)$} & $[H^{\dagger i} \phi^a(\sigma^a)_i^{\ j} H_j]\, B^{\mu \nu} B_{ \mu \nu}$\\[1ex] \cline{2-2}
        \rule{0pt}{3.5ex} & $[H^{\dagger i} \phi^a(\sigma^a)_i^{\ j} H_j]\, G^{\mu \nu} G_{ \mu \nu}$\\[1ex] \cline{2-2}
        \rule{0pt}{3.5ex} & $\phi^a\,[H^{\dagger i} (\sigma^a)_i^{\ j} H_j]\, W^{\mu \nu,b} W^b_{ \mu \nu}$\\[1ex] \cline{2-2}
        \rule{0pt}{3.5ex} & $\phi^a\,[H^{\dagger i} (\sigma^b)_i^{\ j} H_j]\, W^{\mu \nu,a} W^b_{ \mu \nu}$\\[1ex] \cline{2-2}
        \rule{0pt}{3.5ex} & $|H|^2 \phi^{a}\,W^{\mu\nu,a}B_{\mu\nu}$ \\[1ex] \cline{2-2}
        \rule{0pt}{3.5ex}  & \ \ $\varepsilon^{abc}\, \phi^a  [H^{\dagger i} (\sigma^b)_i^{\ j} H_j]\, W^{\mu \nu,c} B_{\mu \nu }$\ \\[1ex] \cline{2-2}
        \rule{0pt}{3.5ex} & $D^{\mu}\phi_{ij }D^{\nu}H^{i}H^{\dagger j}\,B_{\mu\nu}$ \\[1ex] \cline{2-2}
        \rule{0pt}{3.5ex} & $D^{\mu}\phi_{ij }\, H^{\dagger k}D^{\nu}H_{k}\,W^{ij}_{\mu\nu}$ \\[1ex] \cline{2-2}
        \rule{0pt}{3.5ex} & $[H^{\dagger i} H^{ j}  \phi_{ij}]\,|D_\mu H|^2$ \\[1ex] \cline{2-2} 
        \rule{0pt}{3.5ex} & $\phi_{ij}\,|H|^2 D^{\mu}H^{i} D_{ \mu}H^{\dagger j}$ \\[1ex] \cline{2-2}
        \rule{0pt}{3.5ex} & $[H^{\dagger i}  \phi_{ij} D_{ \mu}H^j][H^{\dagger k} D^{\mu}H_{k}]$ \\[1ex] \cline{2-2}
        \rule{0pt}{3.5ex} & $H^{\dagger} \phi^a_{A}(\sigma^a)_i^{\ j} H_j ( D^{\mu}H)_k (D_{ \mu}H^\dagger)^k$ \\[1ex] \cline{2-2}
        \hline
        \rule{0pt}{3.5ex}\multirow{4}{*}[-3.25ex]{$(\boldsymbol{1},\boldsymbol{5},0)$} & $\phi_{ijkl}\,H^{\dagger i} H^j\, W^{kl\mu\nu}B_{\mu\nu}$ \\[1ex] \cline{2-2}
        \rule{0pt}{3.5ex} & $\phi_{ijkl}\,H^{\dagger i} H^j\, W^{km\mu\nu}W^{ml}_{\mu\nu}$ \\[1ex] \cline{2-2}
        \rule{0pt}{3.5ex} & \ \ $[H^{\dagger i} H^{\dagger j}  \phi_{ijkl}]\,D^{\mu}H^{k} D_{ \mu}H^l$\ \ \ \\[1ex]\cline{2-2}
        \rule{0pt}{3.5ex} & $D^{\mu}\phi_{ijkl }D^{\nu}H^{\dagger i}H^{j}\,W^{kl}_{\mu\nu}$ \\[1ex]
        \hline
        \rule{0pt}{3.5ex}$(\boldsymbol{1},\boldsymbol{7},0)$ & \ \ $\phi_{ijklmn}\,H^{\dagger m} H^n\, W^{ij\mu\nu}W^{kl}_{\mu\nu}$\ \ \ \\[1ex]
        \hline
        \bottomrule
        \end{tabular}}
  \caption{Dimension-seven exotic operators that couple boson pairs to color-singlet bSM fields $\phi$ with specified SM quantum numbers. Indices are as shown in Tables \hyperref[operator-table]{1} and \hyperref[operator-table-2]{2}.}
    \label{operator-table-3}
\end{table*}
\renewcommand{\arraystretch}{1.0}      

\begin{table*}[]
    \centering
    Mass dimension 7 [$\times 1/\Lambda^3$]\\[1ex]
    \begin{tabular}{c||c}  \toprule\hline  
    \rule{0pt}{3.5ex} \ \ $\mathrm{SU}(3)_{\text{c}}\times \mathrm{SU}(2)_{\text{L}}\times \mathrm{U}(1)_Y$\ \ \ & Operators \\[1ex]
        \hline
        \hline
        \rule{0pt}{3.5ex} \multirow{3}{*}[-2.0ex]{$(\boldsymbol{8},\boldsymbol{1},0)$} & $(H^\dagger \sigma^a H)\,\phi\, W^{\mu\nu,a}G_{\mu\nu}$ \\[1ex] \cline{2-2}
        \rule{0pt}{3.5ex} & \ \ $(H^\dagger H)\,\phi^A\, B^{\mu\nu}G^A_{\mu\nu}$ \\[1ex] \cline{2-2}
        \rule{0pt}{3.5ex} & $D^{\mu}\phi_{A}D^{\nu}H_{i}H^{i \dagger}\,G^{A}_{\mu\nu}$ \\[1ex]
        \hline
        \rule{0pt}{3.5ex} \multirow{5}{*}[-4.5ex]{$(\boldsymbol{8},\boldsymbol{3},0)$} & $[H^{\dagger} \phi^a_{A}(\sigma^a)_i^{\ j} H_j]\, G_{A}^{\mu \nu} B_{\mu \nu}^{} $ \\[1ex]\cline{2-2} 
        \rule{0pt}{3.5ex}  & \ \ $d^{ABC}\,  [H^{\dagger i} \phi^a_{A}(\sigma^a)_i^{\ j} H_j]\, G_{B}^{\mu \nu} G_{\mu \nu C}$\ \ \ \\[1ex]\cline{2-2} 
        \rule{0pt}{3.5ex}  & \ \ $|H|^2 \phi^{a}_A\,W^{\mu\nu,a}G_{\mu\nu A}$\\[1ex]\cline{2-2}
        \rule{0pt}{3.5ex}  & \ \ $\varepsilon^{abc}\, \phi^a_{A}  [H^{\dagger i} (\sigma^b)_i^{\ j} H_j]\, W^{\mu \nu,c} G_{\mu \nu A}$\ \\[1ex] \cline{2-2}
        \rule{0pt}{3.5ex} & $D_{\mu}\phi_{ij A}D_{\nu}H^{i}H^{j \dagger}\,G^{A}_{\mu\nu}$ \\[1ex]
        \hline      
        \rule{0pt}{3.5ex} $(\boldsymbol{8},\boldsymbol{5},0)$ & $\phi_{ijkl}\,H^{\dagger i} H^j\, W^{kl\mu\nu}G_{\mu\nu}$ \\[1ex]
        \hline
        \rule{0pt}{3.5ex} $(\boldsymbol{10},\boldsymbol{3},0)$ & $ \ \ \varepsilon^{KLM}[H^{\dagger i}\phi_{IJKij}H^j ]\, G_L^{I\mu\nu}G^J_{M\mu\nu}$\ \ \ \\[1ex]
        \hline
        \rule{0pt}{3.5ex} $(\boldsymbol{27},\boldsymbol{3},0)$ & $ [H^{\dagger i}\phi^{KL}_{IJij}H^j ]\, G_K^{I\mu\nu}G^J_{L\mu\nu}$ \\[1ex]
        \hline   
    \bottomrule
\end{tabular}
\caption{Dimension-seven exotic operators that couple boson pairs to color-charged bSM fields $\phi$ with specified SM quantum numbers. Indices are as shown in Tables \hyperref[operator-table]{1}--\hyperref[operator-table-3]{3}.}
    \label{operator-table-4}
\end{table*}
\renewcommand{\arraystretch}{1.0}

All possible representations can be constructed by taking successive products of the fundamental. These higher-dimensional representations may be denoted as symmetric tensors. Totally symmetric tensors of dimension $d$ and rank $r$ have 
\begin{equation}
n=\frac{(d+r-1)!}{r!\,(d-1)!} 
\end{equation}
independent components. For $\mathrm{SU}(2)$, $d=2$, so $n=r+1$. Thus the $n$-dimensional representation is a rank-$(n-1)$ symmetric tensor. As an example, the $\boldsymbol{6}$ of $\mathrm{SU}(2)$ may be represented as a rank-five tensor $\phi_{ijklm}$. 
We can write the covariant derivative acting on a general $\mathrm{SU}(2)_{\text{L}}$ $n$-multiplet $\Phi$ as
\begin{equation}
    [D_\mu\Phi]_\alpha = [\partial_\mu \delta_\alpha^{\ \,\beta} - i g W_\mu^a (\tau_\mathbf{n}^a)_\alpha^{\ \,\beta}]\Phi_\beta,
\end{equation}
where $\tau_\mathbf{n}^a$ are the generators of the $n$-dimensional $\mathrm{SU}(2)_{\text{L}}$ representation. As is typical, the third generator of the group is diagonalized, and the eigenstates are those states with definite electric charge after electroweak symmetry breaking. Thus, it may be worth explicitly stating the relation between these states and the symmetric $(n-1)$-tensors described above. If we label the isospin values as $\{-J,\dots,J-1,J\}$, then we define 
\begin{equation}
    \Phi_{J-k} = \sqrt{{n-1 \choose k}}\ \phi_{\underbrace{1\dots 1}_{n-1-k} \underbrace{2\dots 2}_{k}}\ \ \ \text{for}\ \ \ 0\leq k \leq 2J,
\end{equation}
where the $\phi_{i_1\dots i_{n-1}}$ are totally symmetric. This ensures that \emph{e.g.} $\phi^{\dagger ijkl}\phi_{ijkl}$ is a canonically normalized mass term. Writing all higher representations in terms of symmetric tensors makes the construction of invariant operators much simpler, as it only requires all indices to be contracted. We define the higher representations as having all lower indices, and when required we raise indices with the invariant Levi-Civita symbol. Once the operators are written, the charged-state interaction terms may be extracted via the above relation.

While models with additional  weak doublets or triplets have been studied extensively, fields in higher dimensional representations of $\mathrm{SU}(2)_{\text{L}}$ have received less attention. Of particular interest is the five-dimensional representation. Explicitly, a scalar quintuplet of $\mathrm{SU}(2)_{\text{L}}$ has isospin components $\Phi = (\Phi^{++},\Phi^+,\Phi^0,\Phi^-,\Phi^{--})$.
Here we consider the quintuplet to have zero hypercharge, so the electric charges of the states are integral and range from $-2$ to $+2$. The $\boldsymbol{5}$ of $\mathrm{SU}(2)_{\text{L}}$ is a real representation, so we may enforce $\Phi^{--} = (\Phi^{++})^\dagger$ and $\Phi^{-} = (\Phi^{+})^\dagger$. We express the quintuplet as a rank-four symmetric tensor $\phi_{ijkl}$, and we use group theory to outline the possible singlet operators.

\subsection{Filling out the catalog; phenomenological observations}
\label{s2.3}

With some formalism in hand for higher-dimensional representations of $\mathrm{SU}(2)_{\text{L}}$, we now describe some illustrative examples of operators which will also help demonstrate the use of tensor products to find all singlets. As we wrote above, we have the $\mathrm{SU}(2)$ the tensor product decomposition $\boldsymbol{3} \otimes \boldsymbol{3} = \boldsymbol{1} \oplus \boldsymbol{3} \oplus \boldsymbol{5}$. We may identify the fields in the three-dimensional representation of $\mathrm{SU}(2)_{\text{L}}$ with the gauge bosons, so that the product $\boldsymbol{3} \otimes \boldsymbol{3}$ corresponds to the bilinear $W_{\mu\nu}W^{\mu\nu}$. We can now find a singlet product of LEX states that can marry this bilinear to make a weak singlet.  

The $\boldsymbol{5}$ may correspond to a single LEX state in the five-dimensional representation of SU(2), denoted as above by $\phi_{ijkl}$. Thus the quintuplet may couple to two $W$ field-strength tensors. For illustration, we expand this operator in terms of the charged components of the exotic scalar:
\begin{multline}
    \frac{1}{\Lambda}\,\phi_{ijkl}\, W^{ij \mu\nu}W^{kl}_{\mu\nu}  = \frac{1}{\Lambda}\bigg[\Phi^{++}W^{-\mu\nu}W^-_{\mu\nu} +\Phi^{--}W^{+\mu\nu}W^+_{\mu\nu} - \sqrt{2}\Phi^{+}W^{-\mu\nu}W^3_{\mu\nu} \\ - \sqrt{2}\Phi^{-}W^{+\mu\nu}W^3_{\mu\nu} - \sqrt{\frac{2}{3}}\,\Phi^{0}W^{-\mu\nu}W^+_{\mu\nu} + \sqrt{\frac{2}{3}}\,\Phi^{0}W^{3\mu\nu}W^3_{\mu\nu} \bigg]
\end{multline}
This allows the scalar to decay into boson pairs.
Also of interest are the four- and six-dimensional representations, which must be complex in order for the states to have integral electric charges. Their weak hypercharges must be equal or opposite to that of the SM Higgs, and they may couple to boson pairs via dimension-six invariant operators. Explicitly, the quadruplet has states $\Phi = (\Phi^+, \Phi^0,\Phi^-, \Phi^{--})$ in the $(\boldsymbol{1},\boldsymbol{4},-\frac{1}{2})$ representation of the SM gauge group, and the sextet has states $\Phi = (\Phi^{++},\Phi^{+},\Phi^0, \Phi^-,\Phi^{--}, \Phi^{---})$ in the $(\boldsymbol{1},\boldsymbol{6},-\frac{1}{2})$ representation. 

Moving on, we note that we can compose more singlets by inserting Higgs fields to absorb SU(2) tensor indices. For example, we have the tensor product
\begin{equation}
\boldsymbol{2} \otimes \boldsymbol{6} \supset \boldsymbol{5}\ \ \text{as with spins}\ \  \tfrac{1}{2} \otimes \tfrac{5}{2} = 1 \oplus \underline{2} \oplus 3.
\end{equation}
Identifying the SU(2) doublet with the Higgs and the sextet with the LEX field $\phi_{ijklm}$, we can construct the singlet $\phi_{ijklm}\, H^iW_{\mu\nu}^{jk}W^{lm \mu\nu}$. This corresponds to the iterated tensor product invariant $\boldsymbol{2} \otimes \boldsymbol{6} \otimes \boldsymbol{3} \otimes \boldsymbol{3}$. The tensor product of a quadruplet with a single Higgs doublet, meanwhile, contains a triplet and quintuplet; that is,
\begin{equation}
    \boldsymbol{2} \otimes \boldsymbol{4} = \boldsymbol{3} \oplus \boldsymbol{5},
\end{equation}
where in the latter, the exotic state is ``promoted'' to a $\boldsymbol{5}$, and in the former the state is ``demoted'' to a $\boldsymbol{3}$. Thus, we consider two operators:
\begin{equation}
   \mathcal{L} \supset \frac{1}{\Lambda^2}\, \phi_{ijk}\,H_l \,W^{ij\mu\nu}W^{kl}_{\mu\nu}\ \ \ \mathrm{and}\ \ \ \mathcal{L} \supset \frac{1}{\Lambda^2}\,\phi_{ijk}\,H^k\, W^{ij\mu\nu}B_{\mu\nu}  ,
\end{equation}
which allow the quadruplet to couple to two $\mathrm{SU}(2)_{\text{L}}$ gauge bosons or to one $\mathrm{SU}(2)_{\text{L}}$ and one weak-hypercharge $\mathrm{U}(1)_Y$ gauge boson, respectively. Similarly, we may use alternate tensor products such as
\begin{align}
    \boldsymbol{2} \otimes \boldsymbol{2} \otimes \boldsymbol{3} \supset \boldsymbol{5}\ \ \ \text{and}\ \ \ \boldsymbol{2} \otimes \boldsymbol{2} \otimes \boldsymbol{7} \supset \boldsymbol{5}
\end{align} to form singlets that couple $W$ pairs to the triplet or septet fields at dimensions six or seven.

We note that in addition to coupling directly to pairs of gauge field-strength tensors, LEX states may couple to pairs of weak gauge bosons by coupling to the Higgs kinetic term $D^{\mu}H\, D_{ \mu}H^\dagger$ \cite{Carpenter:2012zr}. For example, the LEX $\mathrm{SU}(2)_{\text{L}}$ triplet appears in the dimension-five operator
\begin{align}
    \mathcal{L} \supset \frac{1}{\Lambda}\,\phi_{ij}\, D^{\mu}H^i\,D_{ \mu}H^{\dagger j}.
    \end{align}
    Once Higgs VEVs are inserted, this becomes an operator of effective dimension three:
\begin{align}
    \mathcal{L} \to \frac{v^2}{\Lambda}\,\phi\, A^{\mu}A_{\mu}.
    \end{align}
    With additional Higgs insertions, LEX states in four- and five-dimensional representation of $\mathrm{SU}(2)_{\text{L}}$ may also appear in this manner for operators up to mass dimension seven.

Finally, operators may have a combination of gauge field-strength tensors and covariant derivatives acting upon Higgs doublets that precipitate the diboson couplings. An interesting example is the dimension-six operator
\begin{align}
    \mathcal{L} \supset \frac{1}{\Lambda^2}\, D^{\mu}\phi_{i A}\,D^{\nu}H^{i}\,G^{A}_{\mu\nu}.
    \end{align}
This operator involves the color-octet weak-doublet scalar. If the Higgs VEV is inserted into this operator, it couples the octet to a gluon and weak gauge boson. However, if the Higgs VEV is not inserted, it has the interesting property of coupling the octet to a gluon-Higgs pair. 

The phenomenology of the diboson portal operators is very interesting and complex. One production process common to all diboson operators is associated production, $pp\to \phi V$, where a LEX field is produced along with one SM gauge boson. Additionally, any non-singlet LEX field may be pair produced via gluon fusion or vector boson fusion. What happens after production of these LEX states can be spectacular. As mentioned above, the LEX multiplets in larger representations of SU(2) contain multiply charged states.  In order to ensure that all operators preserve electromagnetism, the weak hypercharge assignments of LEX states always guarantee integer electric charges of these states.  We then find a characteristic multi-gauge boson topology for the collider production of these states. We have previously studied some models with gauge boson final states \cite{Carpenter:2021tnq,Carpenter:2021gpl}, but there is a lot of ground left to cover. One interesting example is the associated production of a multiply charged $\mathrm{SU}(2)_{\text{L}}$ quintet state via the effective operator
\begin{align}
    \mathcal{L} \supset \frac{1}{\Lambda}\,\phi_{ijkl}\,W^{\mu\nu ij}W^{kl}_{\mu\nu}.
    \end{align}
Here the doubly charged state is produced in association with a $W$; namely, $pp\to\Phi^{++}W^-$, where the charged state decays to same-sign $W$s via $\Phi^{++}\to W^+W^+$. Thus the entire process contains three $W$s, two of which produce a mass resonance, and the other of which may be significantly boosted. The full process can be written as
\begin{align}
    pp\to\phi^{++}W^-\to(W^+W^+)\,W^-,
\end{align}
where the brackets indicate the mass resonance. An even more complex example of this might be the production and decay of a weak-sextet scalar through \emph{e.g.} the dimension-six operator
\begin{align}
    \mathcal{L} \supset \frac{1}{\Lambda^2}\,\phi_{ijklm}H^m \,W^{\mu\nu\,ij}W^{kl}_{\mu\nu}.
    \end{align}
This state contains the fields $\Phi = (\Phi^{+++},\Phi^{++},\Phi^+, \Phi^0,\Phi^-, \Phi^{--})$. We may then have a simple electroweak quark-fusion process like $qq\to \Phi^{+++} \Phi^{--}$. Provided that there is a mass-splitting term for the multiplet, the multiply charged states will cascade decay via the electroweak interaction; for example, $\Phi^{+++}\to \Phi^{++} W^+ \rightarrow \Phi^{+} W^+ W^+$ . The $\Phi^{-}$ $\Phi^{+}$ might decay to $WZ$ through the effective operator. Thus the entire process contains seven gauge bosons:
\begin{align}
qq\to \Phi^{+++} \Phi^{--}\to (W^+(W^+(W^+ Z)))\, (W^- (W^- Z)),
\end{align}
where brackets once more indicate the mass resonances in the decay chains. It happens \cite{Lavoura:1993nq} that such a mass splitting can be introduced via the interaction
\begin{equation}
    \mathcal{L} \supset \lambda (\Phi^\dagger \tau_{\boldsymbol{6}}^a\, \Phi)(H^\dagger \tau_{\boldsymbol{2}}^a H).
\end{equation}
After electroweak symmetry breaking, the Higgs VEV results in differing mass contributions to the different isospin states of the bSM multiplet. The triply charged state would be the heaviest, and this would allow for the above decay chain.
\section{Charge flow and cross sections}
\label{s3}

In this section, we examine the utility of specifying the light-exotic content of a theory through the lens of some simple examples. In particular, we introduce two families of toy models with various bSM color-charged fields and demonstrate how the $\mathrm{SU}(3)_{\text{c}}$ representation alone, with all else being equal, can dramatically affect LHC cross sections. These examples are part of a class totally separate from the LEX scalars communicating with the Standard Model through the diboson portal; we hope that the change of tack serves to demonstrate the breadth of the theory space that can be explored with LEX-EFT.

\subsection{Completing an exotic operator with more exotics}
\label{s3.1}

\begin{figure}\label{UVcompFig}
\begin{align*}
\scalebox{0.75}{\begin{tikzpicture}[baseline={([yshift=-0.9ex]current bounding box.center)},xshift=12cm]
\begin{feynman}[large]
\vertex (i1);
\vertex [right = 1.7cm of i1] (i2);
\vertex [above left=1.5 cm of i1] (v1);
\vertex [below left=1.5cm of i1] (v2);
\vertex [above right=1.5cm of i2] (v3);
\vertex [below right=1.5cm of i2] (v4);
\diagram* {
(i1) -- [ultra thick, fermion] (i2),
(v1) -- [ultra thick, fermion] (i1),
(v2) -- [ultra thick, gluon] (i1),
(i2) -- [ultra thick, charged scalar] (v3),
(i2) -- [ultra thick, fermion] (v4),
};
\end{feynman}
\node at (-1.1,0.52) {$q$};
\node at (-1.1,-0.52) {$g$};
\node at (0.78,0.4) {$\Psi_3$};
\node at (2.8,0.65) {$\varphi$};
\node at (2.9,-0.65) {$q^{\text{c}}$};
\end{tikzpicture}}
\end{align*}
\begin{align*}
    \scalebox{0.75}{\begin{tikzpicture}[baseline={([yshift=-.5ex]current bounding box.center)},xshift=12cm]
\begin{feynman}[large]
\vertex (t1);
\vertex [below=1.5cm of t1] (t2);
\vertex [above left=0.5 cm and 1.1666cm of t1] (i1);
\vertex [right=1.75cm of t1] (f1);
\vertex [above right=0.5cm and 1.1666cm of t1] (p1);
\vertex [below right=0.5cm and 1.1666cm of t2] (p3);
\vertex [below left=0.5cm and 1.1666cm of t2] (i2);
\vertex [right=1.75cm of t2] (f2);
\vertex [below right=0.5 cm and 1.1666cm of f2] (p2);
\vertex [above right=0.5cm and 1.1666cm of f2] (p4);
\diagram* {
(i1) -- [ultra thick, gluon] (t1),
(i2) -- [ultra thick, fermion] (t2),
(t2) -- [ultra thick, charged scalar] (t1),
(t1) -- [ultra thick, charged scalar] (p1),
(t2) -- [ultra thick, fermion] (p3),
};
\end{feynman}
\node at (-1.2,0) {$g$};
\node at (-1.2,-1.6) {$q$};
\node at (-0.45,-0.72) {$\Phi_6$};
\node at (1.2, 0.1) {$\varphi$};
\node at (1.2,-1.55) {$q^{\text{c}}$};
\end{tikzpicture}}
\end{align*}
\begin{align*}
\scalebox{0.75}{\begin{tikzpicture}[baseline={([yshift=-.75ex]current bounding box.center)},xshift=12cm]
\begin{feynman}[large]
\vertex (i1);
\vertex [right = 1.25cm of i1] (i2);
\vertex [above left=1.5 cm of i2] (d1);
\vertex [below left=1.5cm of i2] (d2);
\vertex [right= 1.25cm of i2] (g1);
\vertex [above left = 0.45 cm and 0.2 cm of g1] (v1p);
\vertex [left = 0.875cm of v1p] (fu1);
\vertex [below left = 0.47 cm and 0.2 cm of g1] (v2p);
\vertex [left=0.855cm of v2p] (fu2);
\vertex [above right = 0.625 cm and 0.625 cm of i2] (l1);
\vertex [below right = 0.625 cm and 0.625 cm of i2] (l2);
\vertex [above right=1.5 cm of g1] (v1);
\vertex [below right=1.5cm of g1] (v2);
\diagram* {
(d1) -- [ultra thick, fermion] (fu1),
(d2) -- [ultra thick, gluon ] (fu2),
(fu1) -- [ultra thick,fermion, quarter right, looseness=0.97] (fu2),
(i2) -- [ultra thick, scalar, half left, looseness=1.7] (g1),
(g1) -- [ultra thick, scalar, half left, looseness=1.7] (i2),
(v1p) -- [ultra thick, charged scalar] (v1),
(v2p) -- [ultra thick, fermion] (v2),
(fu2) -- [ultra thick, fermion, quarter right, looseness=0.85] (v2p),
};
\end{feynman}
\node at (0.1,0.7) {$q$};
\node at (0.1,-0.7) {$g$};
\node at (3.7,0.72) {$\varphi$};
\node at (3.7,-0.72) {$q^{\text{c}}$};
\node at (0.95,0) {$q$};
\node at (1.87,-0.97) {$q$};
\node at (2.9,0) {$\Phi_6$};
\node at (1.9,0.91) {$\Phi_8$};
\node[isosceles triangle,
    draw,shape border rotate= 90,
    fill=black,
    inner sep = 1.6pt] (T)at (2.5,-0.05){};
\end{tikzpicture}}
\end{align*}
\caption{Diagrams representing toy UV completions of the operator \eqref{sQQgModel} by way of (top) a heavy quark $\Psi_3$, (middle) a heavy sextet scalar $\Phi_6$, and (bottom, at loop level) a heavy sextet scalar $\Phi_6$ and octet scalar $\Phi_8$.}
\end{figure}
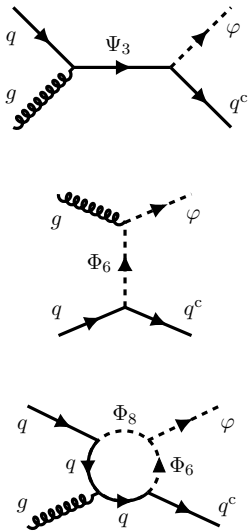

Consider first an operator governing an $\mathrm{SU}(3)_{\text{c}}$ color-sextet scalar $\varphi$ and quarks and gluons but no leptons:
\begin{align}\label{sQQgModel}
\mathcal{L}_{\Phi qqg} \supset \frac{1}{\Lambda^2}\,\lambda_{qq}^{IJ}\,\Pi_s^{\ a\,ij}\,\varphi^{\dagger s}\,(\,\overbar{q^{\text{c}}_{\text{R}}}_{Ii}\,\sigma^{\mu\nu}q_{\text{R}Jj})\, G_{\mu\nu\,a} + \text{H.c.}.
\end{align}
The electric charge of $\varphi$ is $Q = 1/3$ if the two quarks are of unequal type (\emph{i.e.}, one up and one down). The coefficients $\Pi_s^{\ a\,ij}$ are Clebsch-Gordan coefficients that project out a color singlet from the direct-product representation $\boldsymbol{3} \otimes \boldsymbol{3} \otimes \boldsymbol{\bar{6}} \otimes \boldsymbol{8}$. The operator \eqref{sQQgModel} produces multijet events at LHC through single sextet production, and a cursory investigation of a similar operator is carried out in \hyperref[s4]{Section 4}. Here, we simply wish to observe that the group-theoretic coefficients $\Pi_s^{\ a \,ij}$ --- and the size of resulting cross sections --- are not unique. In particular, there is an operator of the form \eqref{sQQgModel} for each independent color singlet that can be formed from the given direct-product color representation. How to proceed depends on one's point of view. On one hand, the number of such independent singlets is finite, and it is possible to construct a single operator whose color factor is expressed as a linear combination of (all of the) independent Clebsch-Gordan coefficients. On the other hand, operators with linearly independent Clebsch-Gordan coefficients are certainly distinct operators, and it may be reasonable --- and phenomenologically interesting --- to investigate an individual infrared operator while \emph{assuming} that its particular Clebsch-Gordan coefficients depend on the representation of the ultraviolet degree(s) of freedom that have been integrated out. To demonstrate this \emph{portal-based} approach, which extends the line of thinking begun in \hyperref[s2]{Section 2}, we provide three toy UV completions of this operator in \hyperref[UVcompFig]{Figure 1}.

The top diagram in \hyperref[UVcompFig]{Figure 1} shows an $s$-channel completion by way of a color-triplet fermion $\Psi_3$. This completion produces the desired color structure by contracting a $\boldsymbol{\bar{3}}$ with a $\boldsymbol{3}$ in
\begin{align}
\boldsymbol{3} \otimes \boldsymbol{\bar{3}} \otimes \boldsymbol{8}\ \ \ \text{and}\ \ \ \boldsymbol{3} \otimes \boldsymbol{3} \otimes \boldsymbol{\bar{6}}.
\end{align}
The vertex corresponding to the first color invariant, which couples a gluon to two different quarks, arises at loop level in minimal extensions of the Standard Model with compactified extra spatial dimensions, which preserve Kaluza-Klein (KK) parity \cite{PhysRevD.66.056006,Freitas:2018bag}, and at tree level in KK parity-violating models \cite{Bhattacherjee:2008kx} (in either case, the role of the color-triplet fermion $\Psi_3$ is played by \emph{e.g.} a level-one KK quark). Explicitly, the coefficients in \eqref{sQQgModel} in this scenario take the form
\begin{align}\label{Pi3}
[\Pi_{\boldsymbol{3}}]_s^{\ a\,ij} = \bt{K}_s^{\ \,ik}[\bt{t}_{\boldsymbol{3}}^a]_k^{\ \,j},
\end{align}
where $\bt{t}_{\boldsymbol{3}}^a$ are the generators of the fundamental representation of $\mathrm{SU}(3)$ and the Clebsch-Gordan coefficients $\bt{K}_s^{\ \,ij}$, which are symmetric in $\mathrm{SU}(3)$ fundamental indices, uniquely project the color singlet out of the direct product $\boldsymbol{3} \otimes \boldsymbol{3} \otimes \boldsymbol{\bar{6}}$ \cite{Han:2009ya}. It happens that the squared norm of the array \eqref{Pi3}, which would for instance be computed as part of the cross section $\sigma(qg \to \varphi q^{\text{c}})$, is given by
\begin{align}\label{squared3}
\bt{K}_s^{\ \,ik}[\bt{t}_{\boldsymbol{3}}^a]_k^{\ \,j}[\bt{t}_{\boldsymbol{3}}^a]_j^{\ \,k'} \bar{\bt{K}}{}^s_{\ ik'} = 8.
\end{align}
The operator \eqref{sQQgModel} could instead be completed by yet another sextet scalar $\Phi_6$, as shown by the middle diagram in \hyperref[UVcompFig]{Figure 1}, in which case a $\boldsymbol{\bar{6}}$ is contracted with a $\boldsymbol{6}$ in
\begin{align}
\boldsymbol{6}\otimes \boldsymbol{\bar{6}} \otimes \boldsymbol{8}\ \ \ \text{and}\ \ \ \boldsymbol{3} \otimes \boldsymbol{3} \otimes \boldsymbol{\bar{6}}.
\end{align}
While exotic, this model can be envisioned as a color-sextet analog of the extra-dimensional model with $\mathrm{SU}(3)_{\text{c}}$ KK excitations considered above. In this case, the group-theoretic factors are
\begin{align}\label{Pi6}
[\Pi_{\boldsymbol{6}}]_s^{\ a\,ij} = \bt{K}_r^{\ \,ij}[\bt{t}_{\boldsymbol{6}}^a]_s^{\ \,r},
\end{align}
where in this simple example the only difference is in the generators $\bt{t}_{\boldsymbol{6}}^a$ of the six-dimensional representation of $\mathrm{SU}(3)$ \cite{Carpenter:2021rkl}. The factor that arises in the computation of cross sections in this case is given by
\begin{align}\label{squared6}
\bt{K}_r^{\ \,ij}[\bt{t}_{\boldsymbol{6}}^a]_s^{\ \,r}[\bt{t}_{\boldsymbol{6}}^a]_{r'}^{\ \ s} \bar{\bt{K}}{}^{r'}_{\ \ ij} = 20.
\end{align}
Finally, the operator \eqref{sQQgModel} can be completed yet again at loop level (and without non-diagonal gauge interactions) by introducing two color-charged degrees of freedom in the sextet and adjoint representations of $\mathrm{SU}(3)_{\text{c}}$, as shown in the bottom diagram in \hyperref[UVcompFig]{Figure 1}. This case is interesting because it does not correspond to a single heavy color-charged degree of freedom, and so the color flow must be tracked more carefully. In particular, the loop is built of the color invariants
\begin{align}
\boldsymbol{3} \otimes \boldsymbol{\bar{3}} \otimes \boldsymbol{8}\ \ \ \text{(twice)},\ \ \ \boldsymbol{6}\otimes \boldsymbol{\bar{6}} \otimes \boldsymbol{8},\ \ \ \text{and}\ \ \ \boldsymbol{3} \otimes \boldsymbol{3} \otimes \boldsymbol{\bar{6}},
\end{align}
and the group-theoretic factors in \eqref{sQQgModel} take the form
\begin{align}\label{PiLoop}
[\Pi_{\text{loop}}]_s^{\ a\,ij} &= \bt{K}_r^{\ \,jl}[\bt{t}_{\boldsymbol{3}}^a]_l^{\ \,k} [\bt{t}_{\boldsymbol{3}}^b]_k^{\ \,i}[\bt{t}_{\boldsymbol{6}}^b]_s^{\ \,r}
\end{align}
with square
\begin{align}\label{squaredLoop}
[\Pi_{\text{loop}}]_s^{\ a\,ij}[\bar{\Pi}_{\text{loop}}]^s_{\ a\,ij} = \frac{20}{9}.
\end{align}
The square of \eqref{PiLoop} is therefore \emph{smaller} than both \eqref{squared3} and \eqref{squared6}. We therefore find that cross sections computed within the framework of the effective operator \eqref{sQQgModel} differ by a factor of up to 9, ignoring all other differences including loop factors, depending on the color representation of the UV degree of freedom.

\subsection{A family of toy models for $hh$ production}
\label{s3.2}

For some variety, consider next a toy model in which some color-charged scalar $\phi$ enjoys a renormalizable coupling to the SM Higgs doublet $H$, so that its dynamics are captured by
\begin{align}\label{HiggsUVmodel}
\mathcal{L}_{\phi} \supset (D_{\mu}\phi)^{\dagger}D^{\mu}\phi + \lambda_{\phi H}\,\phi^{\dagger} \phi\, H^{\dagger} H.
\end{align}
\begin{figure}\label{higgsLoopFig}
\begin{align*}
\scalebox{0.75}{\begin{tikzpicture}[baseline={([yshift=-.75ex]current bounding box.center)},xshift=12cm]
\begin{feynman}[large]
\vertex (i1);
\vertex [right = 1.25cm of i1] (i2);
\vertex [above left=1.5 cm of i2] (d1);
\vertex [below left=1.5cm of i2] (d2);
\vertex [right= 1.25cm of i2] (g1);
\vertex [above left = 0.45 cm and 0.2 cm of g1] (v1p);
\vertex [left = 0.875cm of v1p] (fu1);
\vertex [below left = 0.45 cm and 0.2 cm of g1] (v2p);
\vertex [left=0.875cm of v2p] (fu2);
\vertex [above right = 0.625 cm and 0.625 cm of i2] (l1);
\vertex [below right = 0.625 cm and 0.625 cm of i2] (l2);
\vertex [above right=1.5 cm of g1] (v1);
\vertex [below right=1.5cm of g1] (v2);
\diagram* {
(d1) -- [ultra thick, gluon] (fu1),
(d2) -- [ultra thick, gluon ] (fu2),
(i2) -- [ultra thick, charged scalar, half left, looseness=1.7] (g1),
(g1) -- [ultra thick, charged scalar, half left, looseness=1.7] (i2),
(v1p) -- [ultra thick, scalar] (v1),
(v2p) -- [ultra thick, scalar] (v2),
};
\end{feynman}
\node at (0.1,0.64) {$g$};
\node at (0.1,-0.66) {$g$};
\node at (3.7,0.72) {$h$};
\node at (3.7,-0.72) {$h$};
\node at (1.87,0) {$\phi$};
\node[isosceles triangle,
    draw,shape border rotate= 90,
    fill=black,
    inner sep = 1.6pt] (T)at (2.5,-0.05){};
\node[isosceles triangle,
    draw,shape border rotate= -90,
    fill=black,
    inner sep = 1.6pt] (T)at (1.25,0.05){};
\end{tikzpicture}}\ \ \ +\ \ \ \scalebox{0.75}{\begin{tikzpicture}[baseline={([yshift=-.75ex]current bounding box.center)},xshift=12cm]
\begin{feynman}[large]
\vertex (i1);
\vertex [right = 1.25cm of i1] (i2);
\vertex [above left=1.5 cm of i2] (d1);
\vertex [below left=1.5cm of i2] (d2);
\vertex [right= 1.25cm of i2] (g1);
\vertex [above left = 0.45 cm and 0.2 cm of g1] (v1p);
\vertex [left = 0.875cm of v1p] (fu1);
\vertex [below left = 0.45 cm and 0.2 cm of g1] (v2p);
\vertex [left=0.875cm of v2p] (fu2);
\vertex [above right = 0.625 cm and 0.625 cm of i2] (l1);
\vertex [below right = 0.625 cm and 0.625 cm of i2] (l2);
\vertex [above right=1.5 cm of g1] (v1);
\vertex [below right=1.5cm of g1] (v2);
\diagram* {
(d1) -- [ultra thick, gluon] (fu1),
(d2) -- [ultra thick, gluon] (fu2),
(i2) -- [ultra thick, charged scalar, half left, looseness=1.7] (g1),
(g1) -- [ultra thick, charged scalar, half left, looseness=1.7] (i2),
(g1) -- [ultra thick, scalar] (v1),
(g1) -- [ultra thick, scalar] (v2),
};
\end{feynman}
\node at (0.1,0.64) {$g$};
\node at (0.1,-0.66) {$g$};
\node at (3.78,0.72) {$h$};
\node at (3.78,-0.72) {$h$};
\node at (1.87,0) {$\phi$};
\node[isosceles triangle,
    draw,shape border rotate= -90,
    fill=black,
    inner sep = 1.6pt] (T)at (1.25,0.05){};
\end{tikzpicture}}
\end{align*}
\caption{Representative diagrams containing some color-charged scalar $\phi$ resulting in enhanced rates of dihiggs ($hh$) production at the LHC.}
\end{figure}
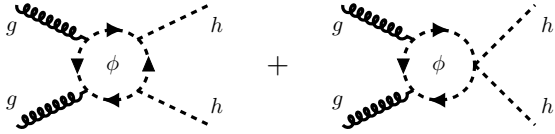This model generates loops of the form displayed in \hyperref[higgsLoopFig]{Figure 2} that could enhance the cross section of Higgs pair production at the LHC, which is encoded in the Wilson coefficient of the dimension-six operator
\begin{align}\label{HiggsIRmodel}
\mathcal{L}_{hh} \supset \frac{1}{\Lambda^2}\,\kappa_{HG}\,H^{\dagger} H \tr G_{\mu\nu}\,G^{\mu\nu}.
\end{align}
Here the trace is over $\mathrm{SU}(3)_{\text{c}}$ adjoint indices. The effect of the exotic scalar $\phi$ on $\sigma(gg \to hh)$ can be significant and depends strongly on the scalar's charge(s). To demonstrate this, we show in \hyperref[gghhFull]{Figure 3} the leading-order (one-loop) LHC dihiggs production cross section as a function of the exotic scalar mass $m_{\phi}$ in two well motivated scenarios: one with a color-triplet scalar, \emph{\`{a} la} squarks, and another with a color-octet scalar. In both scenarios, the scalar-Higgs coupling $\lambda_{\phi H}$ in \eqref{HiggsUVmodel} is set to $0.1$.

\begin{figure}\label{gghhFull}
\centering
\includegraphics[scale=0.75]{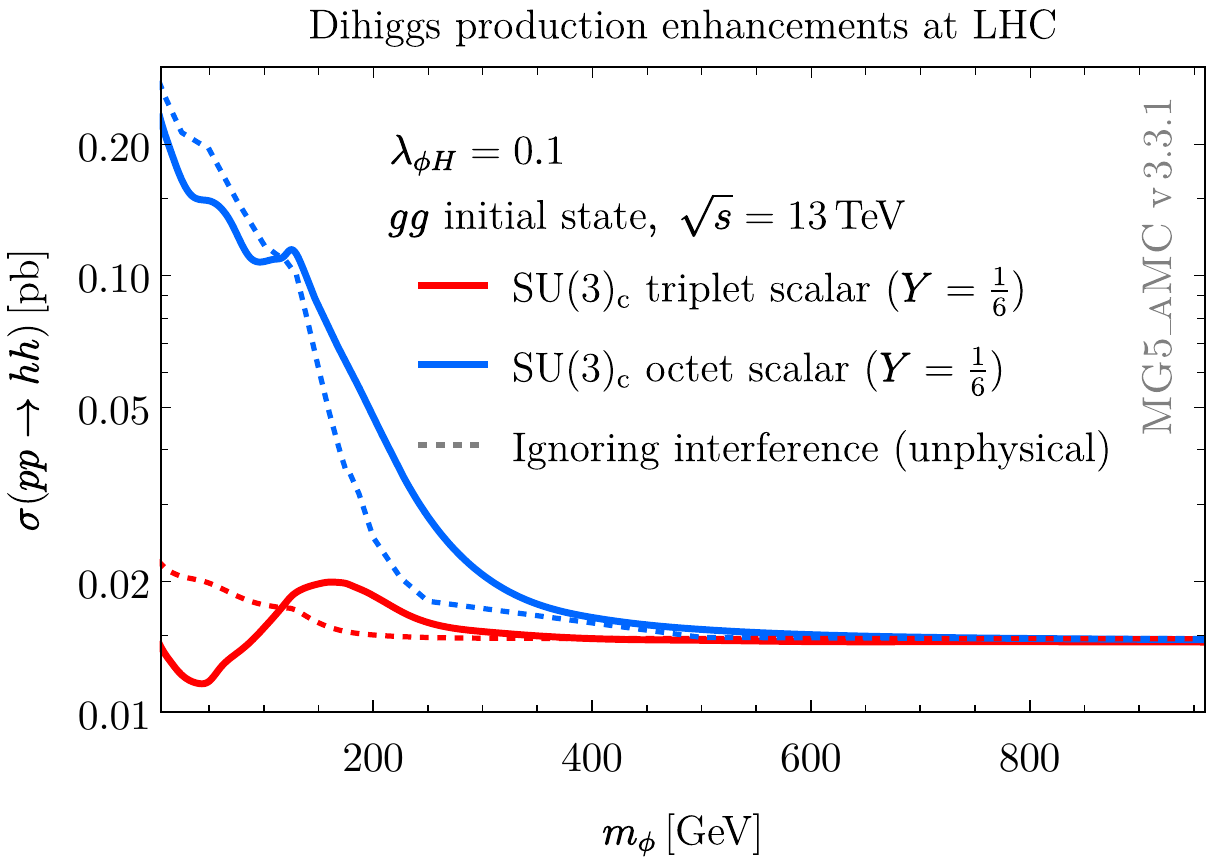}
\caption{Leading-order LHC dihiggs ($hh$) production cross sections at $\sqrt{s}=13\,\text{TeV}$. Leading loops consist of third-generation SM quarks and additional (red) color-triplet or (blue) color-octet scalars $\phi$. Scalar-Higgs coupling $\lambda_{\phi H}$ is set to $0.1$. Only the solid curves correspond to observable cross sections; new-physics contributions are displayed to compare group-theory factors.}
\end{figure}
\begin{figure}\label{ggHHgroup}
\centering
\includegraphics[scale=0.7]{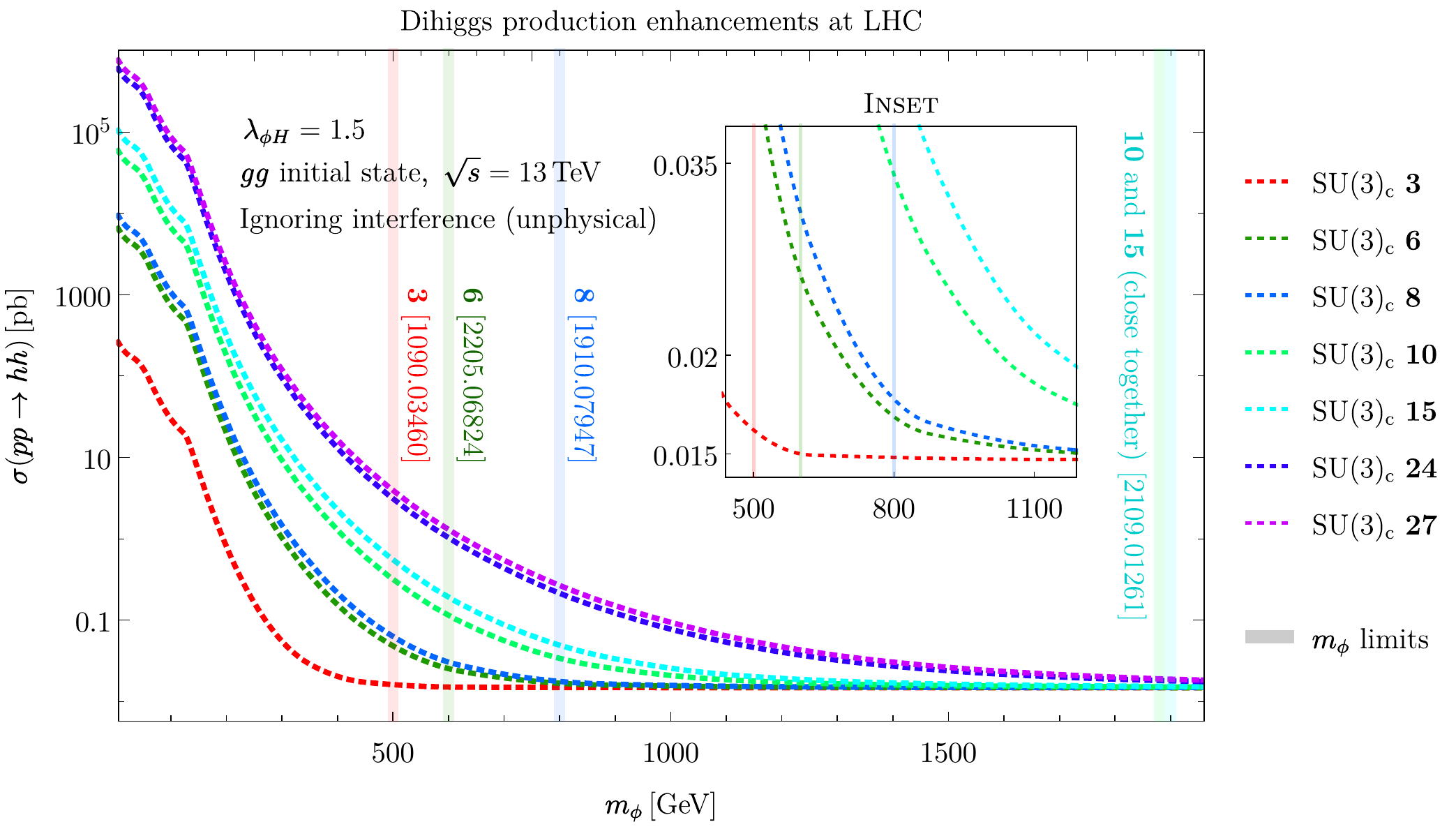}
\caption{Enhancements to leading-order LHC dihiggs ($hh$) production cross sections at $\sqrt{s}=13\,\text{TeV}$, ignoring interference between SM and scalar loops. These new-physics contributions are displayed to compare group-theory factors for increasingly exotic color representations. Scalar-Higgs coupling $\lambda_{\phi H}$ is set to 1.5 to demonstrate scaling (viz. \hyperref[gghhFull]{Figure 3}). Also displayed are robust lower limits on $m_{\phi}$ for selected representations.}
\end{figure}

The enhancement becomes negligible before either scalar reaches $1\,\text{TeV}$ in mass, but in the light-scalar regime where the bSM contribution dominates the SM loops, we find a factor-of-$\mathcal{O}(10)$ difference between the two scenarios. While the interference between SM ($t,b$) loops and $\phi$ loops is intricate, we show using the dashed curves in \hyperref[gghhFull]{Figure 3} that the $\phi$ loops by themselves are responsible for most of this discrepancy.\footnote{For the dashed curves, for visual clarity, we compute $\sigma(gg\to hh)$ in toy models consisting only of $\phi$ loops and add the results to the leading-order SM cross section. This is clearly unphysical, but it gives a qualitative picture of the importance of the group-theoretic factors in the $\phi$ loops.} The responsible group-theoretic term is simply $\tr \bt{t}_{\textbf{r}}^a \bt{t}_{\textbf{r}}^b$, with $\textbf{r}$ the $\mathrm{SU}(3)_{\text{c}}$ representation of $\phi$. It is well known that for the Lie groups $\mathrm{SU}(N)$ this trace is proportional to $\delta^{ab}$, with the factor of proportionality defining the Dynkin index $T_{\textbf{r}}$ of the representation; we provide in \hyperref[DynkinTable]{Table 5} the Dynkin indices of a handful of representations of $\mathrm{SU}(3)$.
\renewcommand\arraystretch{1.0}
\begin{table}\label{DynkinTable}
    \centering
    \begin{tabular}{c | c}
    \toprule
    \hline
     \rule{0pt}{3.5ex}\ \ $\mathrm{SU}(3)_{\text{c}}$ representation \textbf{r}\ \ \ & \ \ \ Dynkin index\ \ \ \ \\[1ex]
     \hline
     \hline
    \rule{0pt}{3.5ex}\ \  $\boldsymbol{3}$\ \ & $\dfrac{1}{2}$\\[1ex]
      \hline
    \rule{0pt}{3.5ex}\ \  $\boldsymbol{6}$\ \ & $\dfrac{5}{2}$\\[1ex]
      \hline
    \rule{0pt}{3.5ex}\ \  $\boldsymbol{8}$\ \ & $3$\\[1ex]
      \hline
    \rule{0pt}{3.5ex}\ \  $\boldsymbol{10}$\ \ & $\dfrac{15}{2}$\\[1ex]
      \hline
    \rule{0pt}{3.5ex}\ \  $\boldsymbol{15}$\ \ & $10$\\[1ex]
      \hline
    \rule{0pt}{3.5ex}\ \  $\boldsymbol{24}$\ \ & $24$\\[1ex]
      \hline
    \rule{0pt}{3.5ex}\ \  $\boldsymbol{27}$\ \ & $27$\\[1ex]
      \hline
      \bottomrule
    \end{tabular}
    \caption{Factors of proportionality (Dynkin indices) in the generator trace $\tr \bt{t}_{\textbf{r}}^a \bt{t}_{\textbf{r}}^b = T_{\textbf{r}}\, \delta^{ab}$ for physically interesting irreducible representations of $\mathrm{SU}(3)$.}
\end{table}
\renewcommand\arraystretch{1}We use these results to extend the $\phi$-only (interference-free) dihiggs enhancements beyond the color-flow limitations of \textsc{MG5\texttt{\textunderscore}aMC} to toy models with $\phi$ in higher representations of $\mathrm{SU}(3)_{\text{c}}$. These rescaled results are displayed in \hyperref[ggHHgroup]{Figure 4} in close analogy with the dashed curves in \hyperref[gghhFull]{Figure 3}. In order to demonstrate the growth of the new-physics enhancements with the scalar-Higgs coupling $\lambda_{\phi H}$, we adopt a different value of $1.5$ in this latest figure. In this benchmark, $\phi$ contributions remain significant into the TeV scale for some color representations: see the inset of \hyperref[ggHHgroup]{Figure 4} for a detailed view of the contributions from low-dimensional representations.

Also visible in \hyperref[ggHHgroup]{Figure 4} are some suggestive, but by no means comprehensive, lower bounds on the mass of $\phi$ in the color representations that have been probed by experiment. This condition applies certainly to the $\boldsymbol{3}$ and $\boldsymbol{8}$ and to lesser extent to the $\boldsymbol{6}$ and $\boldsymbol{10}$ (and higher). We take care to note the intrinsic limitation on these bounds that is imposed by model dependence: it is well known that experimental bounds can shift by $\mathcal{O}(100)\,\text{GeV}$ upon consideration of a new production/decay channel. We therefore highlight in \hyperref[ggHHgroup]{Figure 4} a set of conservative limits that apply to the preponderance of well motivated channels. In particular, we take
\begin{itemize}
    \item $m_{\phi} > 500\,\text{GeV}$ for a color-triplet scalar, still allowed by Run 2 searches for pair-produced squarks $\tilde{q}$ decaying to $q\tilde{\chi}^0$ with relatively heavy $\tilde{\chi}^0$ \cite{ATLAS:2020syg};
    \item $m_{\phi} > 600\,\text{GeV}$ for a color-sextet scalar, allowed for scalars coupling to same-sign quark pairs with reasonable $\mathcal{O}(0.1)$ coupling strengths \cite{Carpenter:2022lhj};
    \item $m_{\phi} > 800\,\text{GeV}$ for a color-octet scalar, the most conservative bound derived fairly recently \cite{Miralles:2019uzg} for CP-even color octet in the Manohar-Wise model \cite{Manohar:2006ga};
    \item $m_{\phi} \gtrsim 1900\,\text{GeV}$ for color-decuplet/quindecuplet scalars that may appear as $R$ hadrons at the LHC in the absence of an efficient decay channel \cite{Preuss:2021ebd}.
\end{itemize}
\hyperref[ggHHgroup]{Figure 4} shows that these limits --- which we reiterate can be strengthened at the expense of model independence --- can be used to probe large scalar-Higgs couplings $\lambda_{\phi H}$ of $\mathcal{O}(1)$. By the same token, we conclude that light --- even sub-TeV --- color-charged scalars are still viable in this $\lambda_{\phi H}$ regime.

\subsection{Group theory impacts on EFT validity and collider reach}
\label{s3.3}

Before we move on, we make some observations about the impacts of charge flow on the range of validity of an effective field theory and, on a related note, the self-consistent experimentally accessible parameter space of such theories. Just above (\emph{viz}. \hyperref[gghhFull]{Figure 3}), we offered a simple example of two models identical in all aspects except for the color charge of the light exotic field $\phi$. The dihiggs production cross sections in these models suggest that the characteristic scale of the EFT obtained by integrating out $\phi$ may vary by $\mathcal{O}(100)\,\text{GeV}$. This line of thinking can be extended to the EFT cutoff $\Lambda$, even though one might suppose that the range of validity of an effective field theory should be unaffected by non-kinematic factors.

We make this notion more concrete by computing the perturbative unitarity bound \cite{PhysRevD.16.1519,Cohen:2021gdw} on the cutoff $\Lambda$ of the operator \eqref{sQQgModel}, which permits processes of the form $qg \to \varphi q^{\text{c}}$. This particular process has angular momentum $J=1/2$. In the massless-quark limit, the perturbative unitarity bound on $\Lambda$ derived by computing the definite-helicity transition amplitude $\mathcal{M}(qg \to \varphi q^{\text{c}})$ is \cite{Carpenter:2022qsw} 
\begin{align}\label{masslessLeptonBound}
\Lambda \geq (\Pi_s^{\ a\,ij}\bar{\Pi}{}^s_{\ a\,ij})^{1/4}\left(\frac{\lambda_{qq}^{IJ}}{2\pi}\right)^{1/2}(\hat{s} - m_{\varphi}^2)^{1/2}
\end{align}
with $m_{\varphi}$ the mass of the light sextet $\varphi$ and the group-theoretic factor $\Pi_s^{\ a\,ij}\bar{\Pi}{}^s_{\ a\,ij}$ given by \eqref{squared3}, \eqref{squared6}, and \eqref{squaredLoop} in the three cases studied in \hyperref[s3.1]{Section 3.1}. Since we found there that the numerical factor could differ by a factor of up to 9, we see from \eqref{masslessLeptonBound} that the perturbative unitarity bound on the operator \eqref{sQQgModel} can vary by a factor as large as
\begin{align}
\left(\frac{[\Pi_{\boldsymbol{6}}]_s^{\ a\,ij}[\bar{\Pi}_{\boldsymbol{6}}]^s_{\ a\,ij}}{[\Pi_{\text{loop}}]_s^{\ a\,ij}[\bar{\Pi}_{\text{loop}}]^s_{\ a\,ij}}\right)^{1/4}= 9^{1/4} \approx 1.73
\end{align}
based purely on the $\mathrm{SU}(3)_{\text{c}}$ representation of the intermediate exotic field (\emph{viz}. \hyperref[UVcompFig]{Figure 1}). This significant effect, which we reemphasize is completely independent of the kinematics of the physical process generated by the operator, has been observed in only a few disparate contexts \cite{DiLuzio:2016sur,Cahill-Rowley:2015aea} --- as far as we are aware --- and deserves greater appreciation, since it can void potentially wide swaths of EFT parameter space on self-consistency grounds.

By the same token, though, it may be that higher minimum EFT cutoffs due to the above mechanism are neutralized by larger cross sections, resulting in net gains in (potentially) accessible parameter space, provided that the effective operator is of high enough mass dimension. The previous example is illustrative: if the cross section $\sigma(qg \to \varphi q^{\text{c}})$ rises by a factor of 9, but the unitarity-bounded cutoff rises by a factor of $\sqrt{3}$, then in principle the experimental reach along the $\Lambda$ axis of EFT parameter space is greater for any given $m_{\varphi}$ despite the higher minimum cutoff. Altogether, therefore, we conclude that even $\mathcal{O}(1)$ numerical factors affect both cross sections and unitarity bounds in effective field theories, ultimately determining how much parameter space should be considered valid and accessible. This observation adds yet another motivation for theorists to comprehensively explore the space of light-exotics models.
\section{Unique kinematics of light-exotics processes}
\label{s4}

We now provide some examples of unique kinematic features that may appear in processes at the LHC and that can only be captured theoretically by specifying the intermediate states that may be (nearly) on shell in such processes. We first consider a pair of models that both produce final states with two hard jets and a pair of charged leptons ($jj\,\ell^+\ell^-$). We then highlight two models that produce the (related) final states with hard jets and missing transverse energy ($jj + E_{\text{T}}^{\text{miss}}$).

\subsection{$jj\,\ell^+\ell^-$: sextet vs. leptoquark}
\label{s4.1}

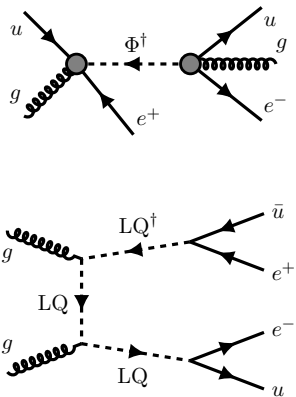
\begin{figure}
\begin{align*}
\scalebox{0.75}{\begin{tikzpicture}[baseline={([yshift=-.75ex]current bounding box.center)},xshift=12cm]
\begin{feynman}[large]
\vertex (i1);
\vertex [right = 2cm of i1] (i2);
\vertex [above right = 1 cm and 1.25 cm of i2] (v1p);
\vertex [right = 1.5 cm of i2] (v2p);
\vertex [below right = 1 cm and 1.25 cm of i2] (f1);
\vertex [above left= 1.3 cm of i1] (p1);
\vertex [below left = 1.3 cm of i1] (p2);
\vertex [below right = 1.25 cm and 1 cm of i1] (l1);
\diagram* {
(i2) -- [ultra thick, charged scalar] (i1),
(i2) -- [ultra thick, fermion] (v1p),
(i2) -- [ultra thick, fermion] (f1),
(v2p) -- [ultra thick, gluon] (i2),
(p1) -- [ultra thick, fermion] (i1),
(p2) -- [ultra thick, gluon] (i1),
(l1) -- [ultra thick, fermion] (i1)
};
\end{feynman}
\node at (-1.07,-0.57) {$g$};
\node at (-1.05,0.6) {$u$};
\node at (1.27,-0.93) {$e^+$};
\node at (1.05,0.38) {$\Phi^{\dagger}$};
\node at (3.5,-0.7) {$e^-$};
\node at (3.4,0.75) {$u$};
\node at (3.6,0.33) {$g$};
\node at (0,0) [circle,draw=black,line width = 0.5mm, fill=gray,inner sep=3.5pt]{};
\node at (2,0) [circle,draw=black,line width = 0.5mm, fill=gray,inner sep=3.5pt]{};
\end{tikzpicture}}
\end{align*}
    \begin{align*}
        \scalebox{0.75}{\begin{tikzpicture}[baseline={([yshift=-.5ex]current bounding box.center)},xshift=12cm]
\begin{feynman}[large]
\vertex (t1);
\vertex [below=1.5cm of t1] (t2);
\vertex [above left=0.5 cm and 1.3cm of t1] (i1);
\vertex [above right=0.3cm and 1.9cm of t1] (f1);
\vertex [above right=0.5cm and 1.3cm of f1] (p1);
\vertex [below right=0.5cm and 1.3cm of f1] (p3);
\vertex [below left=0.5cm and 1.3cm of t2] (i2);
\vertex [below right = 0.3cm and 1.9cm of t2] (f2);
\vertex [below right=0.5 cm and 1.3cm of f2] (p2);
\vertex [above right=0.5cm and 1.3cm of f2] (p4);
\diagram* {
(i1) -- [ultra thick, gluon] (t1) -- [ultra thick, charged scalar] (t2),
(i2) -- [ultra thick, gluon] (t2),
(f1) -- [ultra thick, charged scalar] (t1),
(t2) -- [ultra thick, charged scalar] (f2),
(p3) -- [ultra thick, fermion] (f1),
(p1) -- [ultra thick, fermion] (f1),
(f2) -- [ultra thick, fermion] (p4),
(f2) -- [ultra thick, fermion] (p2),
};
\end{feynman}
\node at (-1.3,0.065) {$g$};
\node at (1.0,0.6) {$\text{LQ}^{\dagger}$};
\node at (-1.3,-1.55) {$g$};
\node at (-0.5,-0.8) {$\text{LQ}$};
\node at (0.9,-2.1) {$\text{LQ}$};
\node at (3.45, 0.84) {$\bar{u}$};
\node at (3.55, -0.22) {$e^+$};
\node at (3.55, -1.2) {$e^-$};
\node at (3.45, -2.35) {$u$};
\end{tikzpicture}}
    \end{align*}
    \caption{Diagrams for (top) $\Phi^{\dagger}$ production with charged leptons, with blobs denoting dimension-six effective vertices, and (bottom) first-generation leptoquark pair production at LHC.}
    \label{LQpairFig}
\end{figure}

Scalar leptoquarks (LQs) \cite{BUCHMULLER1987442}, novel spin-zero $\mathrm{SU}(3)_{\text{c}}$ triplets carrying both baryon and lepton number, have a long history both in unified theories \cite{PhysRevD.8.1240,PhysRevLett.32.438,doi:10.1142/S0217732392000070} and in phenomenological models that accommodate lepton flavor universality violation \cite{Dumont:2016xpj,Angelescu:2021lln,Belanger:2021smw}. A so-called \emph{first-generation} scalar leptoquark $\Phi_{\text{LQ}}$ with electric charge $Q = -1/3$ minimally couples at mass dimension four to electrons and up quarks according to
\begin{align}\label{LQLag}
    \mathcal{L}_{\text{LQ}} \supset \Phi_{\text{LQ}}^{\dagger i} \left[y_{\text{L}}\, \overbar{U_{ia}^{\text{c}}}\, \text{P}_{\text{L}}\, \varepsilon^{ab} E_b + y_{\text{R}}\, \overbar{u_{i}^{\text{c}}}\, \text{P}_{\text{R}} e\right] + \text{H.c.},
\end{align}
where $\{U,E\}$ are the first-generation quark and lepton $\mathrm{SU}(2)_{\text{L}}$ doublets with indices $a,b \in \{1,2\}$, and $\{u,e\}$ are the corresponding weak singlets. The Yukawa-type couplings $y_{\text{L}},y_{\text{R}}$ are considered independent in this analysis. The quantum numbers of this leptoquark are specified in \hyperref[qNumTable]{Table 6}.
\renewcommand\arraystretch{1}
\begin{table}
    \centering
    \begin{tabular}{l|| c | c | c}
    \toprule
    \hline
     \rule{0pt}{3.5ex}\ \ Field\ \ \ & \ \ \ \ \ {$\mathrm{SU}(3)_{\text{c}} \times \mathrm{SU}(2)_{\text{L}} \times \mathrm{U}(1)_Y$}\ \ \ \ &  \ \ \ \,{$B$}\ \ \ \ & \ \ \ \,{$L$}\ \ \ \ \\[1ex]
     \hline
     \hline
    \rule{0pt}{3.5ex}\ \  $\Phi$\ \ & $(\boldsymbol{6},\boldsymbol{1},\tfrac{1}{3})$ & 0 & $-1$\\[1ex]
      \hline
    \rule{0pt}{3.5ex}\ \  $\Phi_{\text{LQ}}$\ \ & $(\boldsymbol{3},\boldsymbol{1},-\tfrac{1}{3})$ & $\tfrac{1}{3}$ & $1$\\[1ex]
      \hline
      \bottomrule
    \end{tabular}
    \caption{Quantum numbers of new scalars in color-sextet scalar ($\Phi$) and first-generation scalar leptoquark ($\Phi_{\text{LQ}}$) models, both of which produce $jj\,\ell^+\ell^-$ final states.}
    \label{qNumTable}
\end{table}
\renewcommand\arraystretch{1}In this model, the relevant LHC production process for the $jj\,\ell^+\ell^-$ channel is QCD pair production,  $gg \rightarrow \Phi_{\text{LQ}}^{\dagger} \Phi_{\text{LQ}}$, followed by the decay $\Phi_{\text{LQ}}\rightarrow ue^-$ and its conjugate. A representative diagram for this process is shown in the lower panel of \hyperref[LQpairFig]{Figure 5}. The CMS Collaboration conducted a search \cite{CMS-EXO-17-009} for this specific process using $\mathcal{L} = 35.9\,\text{fb}^{-1}$ of Run 2 data and, in the absence of a signal, excluded first-generation leptoquarks with masses $m_{\text{LQ}} < 1435\,\text{GeV}$ at 95\% confidence level (CL) \cite{Read:2002cls}.

The same final state can be produced at the LHC in a cousin of the operator \eqref{sQQgModel} containing the following dimension-six interaction(s) between SM fermions and a $Q = 1/3$ color-sextet scalar $\Phi$ \cite{Chivukula_91,Celikel_98,Han:2009ya,PhysRevD.79.054002,Han:2010rf,Carpenter:2022lhj,Carpenter:2022qsw}:
\begin{align}\label{sSmodel}
    \mathcal{L}_{\Phi \ell^-} \supset 
    \frac{1}{\Lambda^2}\,\lambda^{IX}_{u\ell} \bt{J}^{\,s\, ia}\,\Phi_s\,(\,\overbar{u^{\text{c}}_{\text{R}}}_{Ii}\,\sigma^{\mu\nu}\ell_{\text{R}X})\, G_{\mu\nu\,a}
    + \text{H.c.}.
\end{align}
The couplings $\lambda_{u\ell}$ are elements of a matrix in quark and lepton generation space, with $I$ or $X=3$ labeling the heavy generation(s). The coefficients $\bt{J}$ are the generalized Clebsch-Gordan coefficients \cite{Han:2009ya} required to construct gauge-invariant contractions of the direct-product representation $\boldsymbol{3} \otimes \boldsymbol{6} \otimes \boldsymbol{8}$ in $\mathrm{SU}(3)$ \cite{Carpenter:2021rkl}. Here and below, the $\boldsymbol{6}$ is indexed by $s,r,\dots$ and the $\boldsymbol{3}$ by $i,j,\dots$. The quantum numbers of this color-sextet scalar are specified in \hyperref[qNumTable]{Table 6}. Here the relevant LHC process is single sextet production with an associated lepton (for first-generation SM fermions, $ug \to \Phi^{\dagger} e^+$) and subsequent decay through the same operator, $\Phi^{\dagger} \rightarrow uge^-$.\footnote{We highlight antisextet production because the conjugate process has lower initial parton luminosity at LHC, but both processes take place.} The sole diagram for this process is displayed in the upper panel of \hyperref[LQpairFig]{Figure 5}.

\begin{figure}\label{sextetLQcompFig}
\centering
\hspace*{-0.5ex}\includegraphics[scale=0.6888]{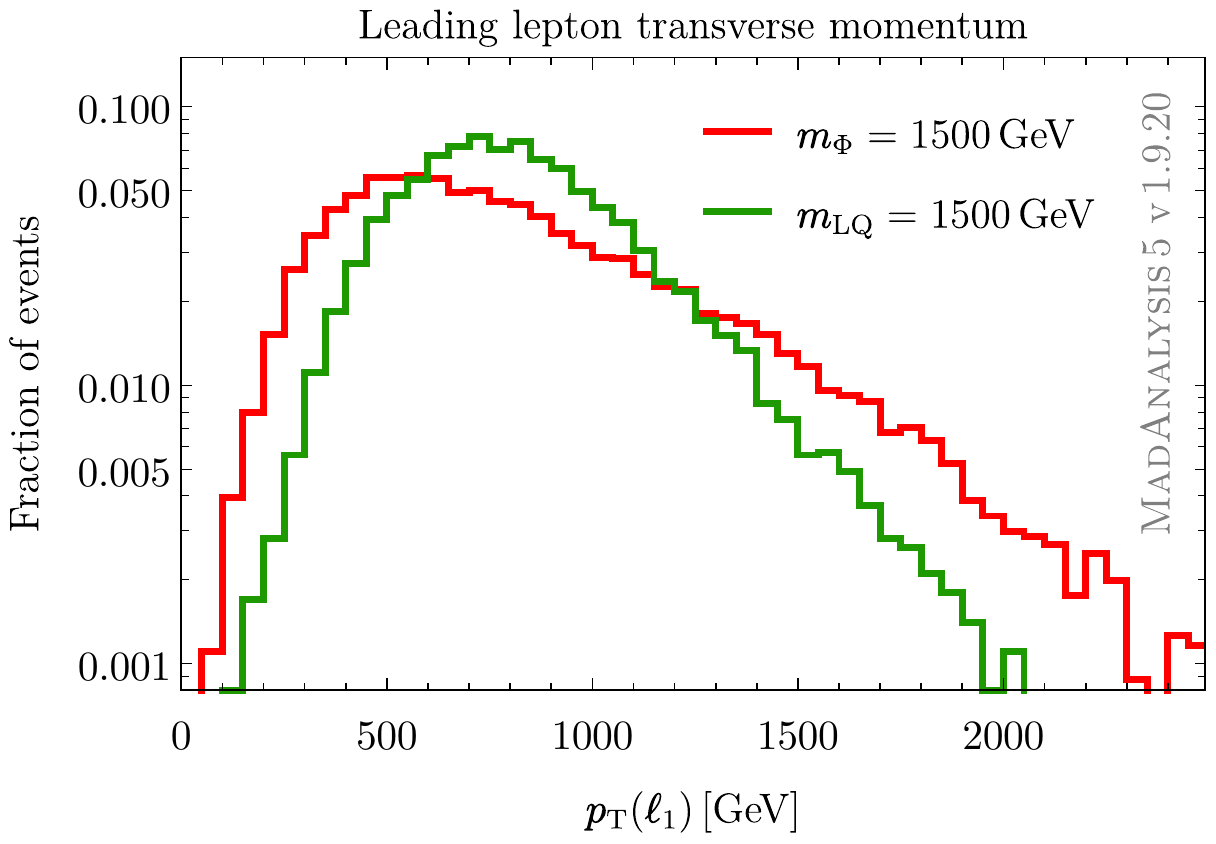}\\
\bigskip
\includegraphics[scale=0.672]{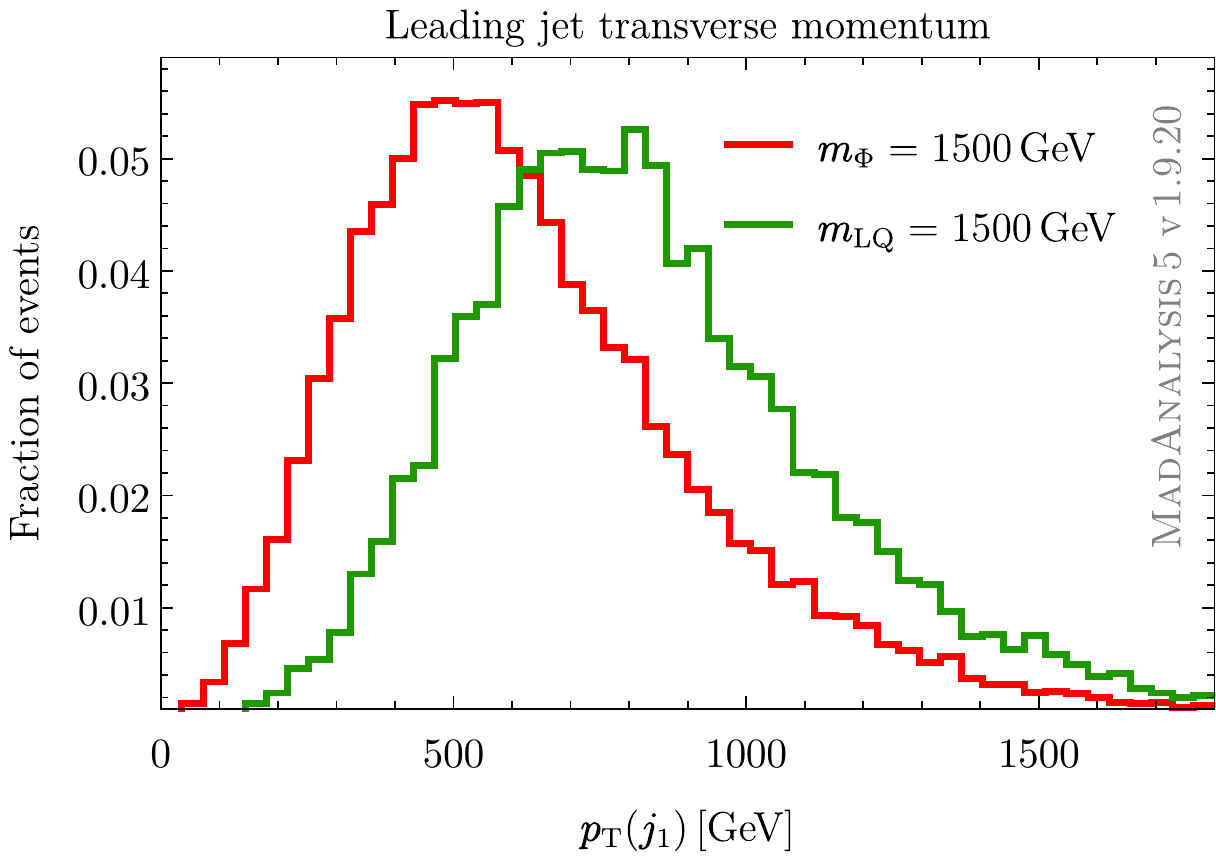}\\
\bigskip
\includegraphics[scale=0.672]{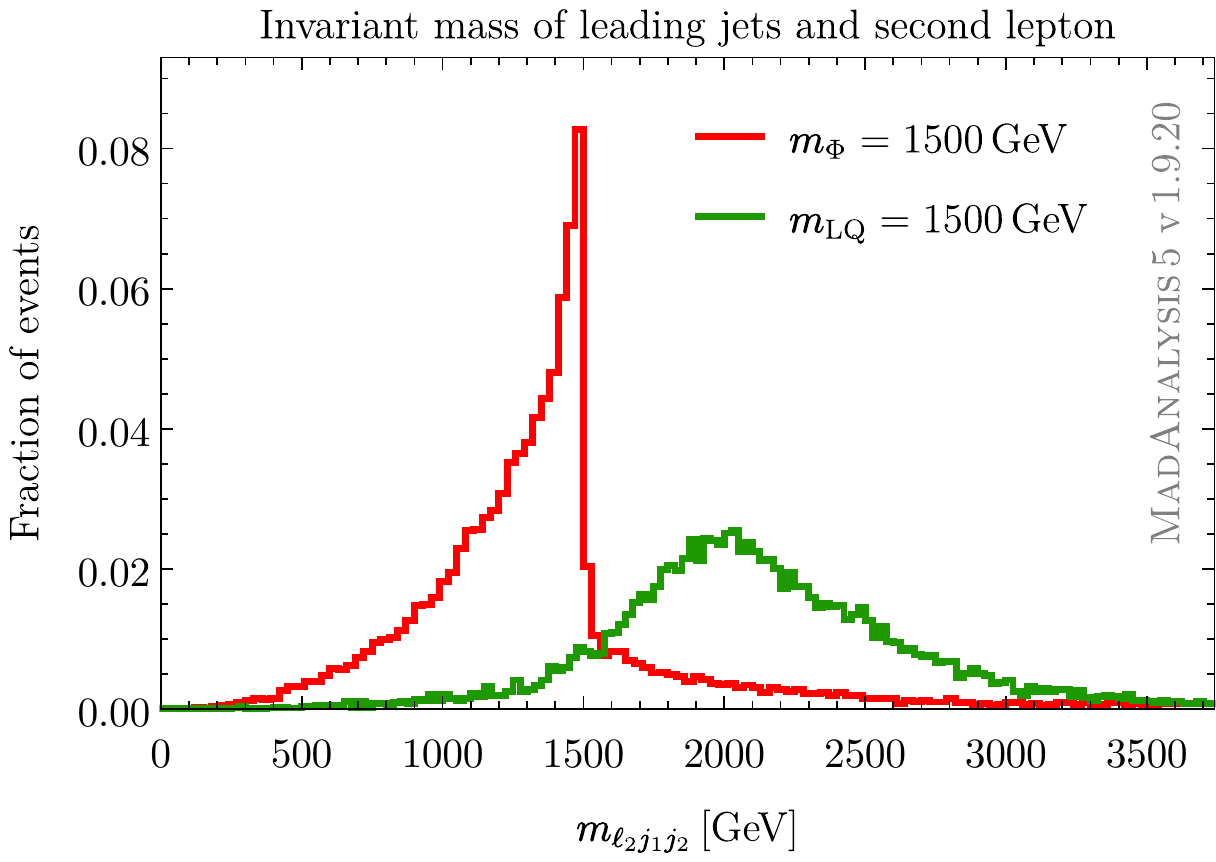}
\caption{Distributions of (top) hardest lepton $p_{\text{T}}$, (center) hardest jet $p_{\text{T}}$, and (bottom) invariant mass of hardest jets and second-hardest lepton for (red) dimension-six sextet scalar production and (green) scalar leptoquark pair production at LHC.}
\end{figure}

In simple scenarios where both exotic scalars couple only to first-generation SM fermions, the final states of these two processes are indistinguishable at the LHC. But the kinematics of these processes are quite different. Some illustrative distributions are compared in \hyperref[sextetLQcompFig]{Figure 6} for exotic scalars set to a common mass of $1.5\,\text{TeV}$, this mass having been chosen in view of the $1.44\,\text{TeV}$ limit on $m_{\text{LQ}}$ mentioned above. The simulated event samples were produced in \textsc{MadGraph5\texttt{\textunderscore}aMC@NLO} (\textsc{MG5\texttt{\textunderscore}aMC}) version 3.3.1 \cite{MG5,MG5_EW_NLO}, showered and hadronized using \textsc{Pythia\,8} version 8.244 \cite{Pythia}, and analyzed with \textsc{MadAnalysis\,5} version 1.9.20 \cite{Conte_2013,Conte_2014,Conte_2018} after performing object reconstruction using its inbuilt simplified fast detector simulator (SFS) \cite{Araz_2021} and an interface to \textsc{FastJet} version 3.3.3 \cite{FJ}. Jets were reconstructed according to the anti-$k_t$ algorithm \cite{Cacciari:2008gp} with radius parameter set to $R=0.4$.

In the top panel of \hyperref[sextetLQcompFig]{Figure 6}, we show that the transverse momentum ($p_{\text{T}}$) of the leading lepton is expected to be significantly higher in the sextet model than for LQ pair production. This is because in the former model, the leading lepton is the one that recoils off of the sextet when it is produced, whereas in the latter model it is a product of one of the decaying leptoquarks. The middle panel shows the leading jet $p_{\text{T}}$, which is likewise expected to be higher in the sextet model but more sharply peaked for LQ pairs. The bottom panel shows the invariant mass $m_{\ell_2 j_1j_2}$ of the two hardest jets and the second lepton. In the sextet model, the system $\ell_2 j_1j_2$ corresponds to the decay products of the exotic scalar, and so the invariant mass can be used to reconstruct the sextet \cite{Carpenter:2022qsw}. No such identification can be made within the leptoquark model, and indeed the distribution is much broader and certainly not peaked at the LQ mass.

\subsection{$jj + E_{\text{T}}^{\text{miss}}$: sextet vs. squark}
\label{s4.2}

\begin{figure}
\begin{align*}
\scalebox{0.75}{\begin{tikzpicture}[baseline={([yshift=-.75ex]current bounding box.center)},xshift=12cm]
\begin{feynman}[large]
\vertex (i1);
\vertex [right = 2cm of i1] (i2);
\vertex [above right = 1 cm and 1.25 cm of i2] (v1p);
\vertex [right = 1.5 cm of i2] (v2p);
\vertex [below right = 1 cm and 1.25 cm of i2] (f1);
\vertex [above left= 1.3 cm of i1] (p1);
\vertex [below left = 1.3 cm of i1] (p2);
\vertex [below right = 1.25 cm and 1 cm of i1] (l1);
\diagram* {
(i2) -- [ultra thick, charged scalar] (i1),
(i2) -- [ultra thick, fermion] (v1p),
(i2) -- [ultra thick, fermion] (f1),
(v2p) -- [ultra thick, gluon] (i2),
(p1) -- [ultra thick, fermion] (i1),
(p2) -- [ultra thick, gluon] (i1),
(l1) -- [ultra thick, fermion] (i1)
};
\end{feynman}
\node at (-1.07,-0.57) {$g$};
\node at (-1.05,0.56) {$d$};
\node at (1.27,-0.97) {$\bar{\nu}_e$};
\node at (1.05,0.38) {$\Phi^{\dagger}$};
\node at (3.5,-0.75) {$\nu_e$};
\node at (3.4,0.75) {$d$};
\node at (3.6,0.33) {$g$};
\node at (0,0) [circle,draw=black,line width = 0.5mm, fill=gray,inner sep=3.5pt]{};
\node at (2,0) [circle,draw=black,line width = 0.5mm, fill=gray,inner sep=3.5pt]{};
\end{tikzpicture}}
\end{align*}
    \begin{align*}
        \scalebox{0.75}{\begin{tikzpicture}[baseline={([yshift=-.5ex]current bounding box.center)},xshift=12cm]
\begin{feynman}[large]
\vertex (t1);
\vertex [below=1.5cm of t1] (t2);
\vertex [above left=0.5 cm and 1.3cm of t1] (i1);
\vertex [above right=0.3cm and 1.9cm of t1] (f1);
\vertex [above right=0.5cm and 1.3cm of f1] (p1);
\vertex [below right=0.5cm and 1.3cm of f1] (p3);
\vertex [below left=0.5cm and 1.3cm of t2] (i2);
\vertex [below right = 0.3cm and 1.9cm of t2] (f2);
\vertex [below right=0.5 cm and 1.3cm of f2] (p2);
\vertex [above right=0.5cm and 1.3cm of f2] (p4);
\diagram* {
(i1) -- [ultra thick, gluon] (t1) -- [ultra thick, charged scalar] (t2),
(i2) -- [ultra thick, gluon] (t2),
(f1) -- [ultra thick, charged scalar] (t1),
(t2) -- [ultra thick, charged scalar] (f2),
(p3) -- [ultra thick, fermion] (f1),
(p1) -- [ultra thick, fermion] (f1),
(f2) -- [ultra thick, fermion] (p4),
(f2) -- [ultra thick, fermion] (p2),
};
\end{feynman}
\node at (-1.3,0.065) {$g$};
\node at (1.0,0.6) {$\tilde{u}^{\dagger}$};
\node at (-1.3,-1.55) {$g$};
\node at (-0.35,-0.7) {$\tilde{u}$};
\node at (0.9,-2.1) {$\tilde{u}$};
\node at (3.45, 0.84) {$\bar{u}$};
\node at (3.55, -0.22) {$\tilde{\chi}^0$};
\node at (3.55, -1.2) {$\tilde{\chi}^0$};
\node at (3.45, -2.35) {$u$};
\end{tikzpicture}}
    \end{align*}
    \caption{Diagrams for (top) $\Phi^{\dagger}$ production with neutrinos, with blobs denoting dimension-six effective vertices, and (bottom) MSSM up squark pair production at LHC. Neutralinos $\tilde{\chi}^0$ appear as missing transverse energy.}
    \label{neutrinoFig}
\end{figure}
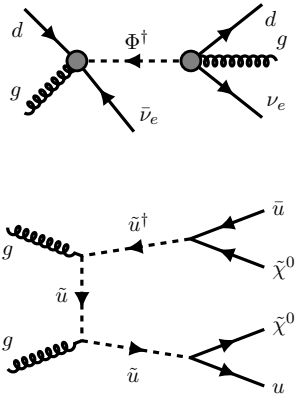

Final states with jets and missing transverse energy have long provided the quintessential search channel for supersymmetry at the LHC since squarks are copiously pair produced in many models (for example, $q\bar{q} \to \tilde{q}^{\dagger} \tilde{q}$) and usually decay to quarks and neutralinos ($\tilde{q} \to q \bar{\tilde{\chi}}^0$ and the conjugate). This is certainly the case in the Minimal Supersymmetric Standard Model (MSSM), which we focus on for simplicity in this work. A representative diagram for this process is displayed in \hyperref[neutrinoFig]{Figure 7}. Meanwhile, an extension of the sextet scalar model \eqref{sSmodel} can produce similar final states at the LHC. Suppose, in particular, that the quarks and leptons are left handed and interact with $\Phi$ according to
\begin{align}\label{sextetNeutrinoModel}
    \mathcal{L}_{\Phi \nu} \supset
    \frac{1}{\Lambda^2}\,\lambda^{IX}_{d\nu}\, \bt{J}^{\,s\, ia}\,\Phi_s\,(\,\overbar{Q^{\text{c}}_{\text{L}}}_{Ii}\doot\sigma^{\mu\nu}L_{\text{L}X})\,G_{\mu\nu\,a} + \text{H.c.}
\end{align}
with notation similar to \eqref{sSmodel}. Then, instead of single sextet production in association with a charged lepton, we have an associated neutrino ($dg \to \Phi^{\dagger} \bar{\nu}$) and the decay(s) $\Phi^{\dagger} \to dg\nu$. These processes notably involve down-type quarks, but aside from the differing SM fermions, the relevant diagram --- displayed in \hyperref[neutrinoFig]{Figure 7} --- is identical to the one in \hyperref[LQpairFig]{Figure 5}.

\begin{figure}\label{sextetMSSMcompFig}
\centering
\hspace*{-0.8ex}\includegraphics[scale=0.682]{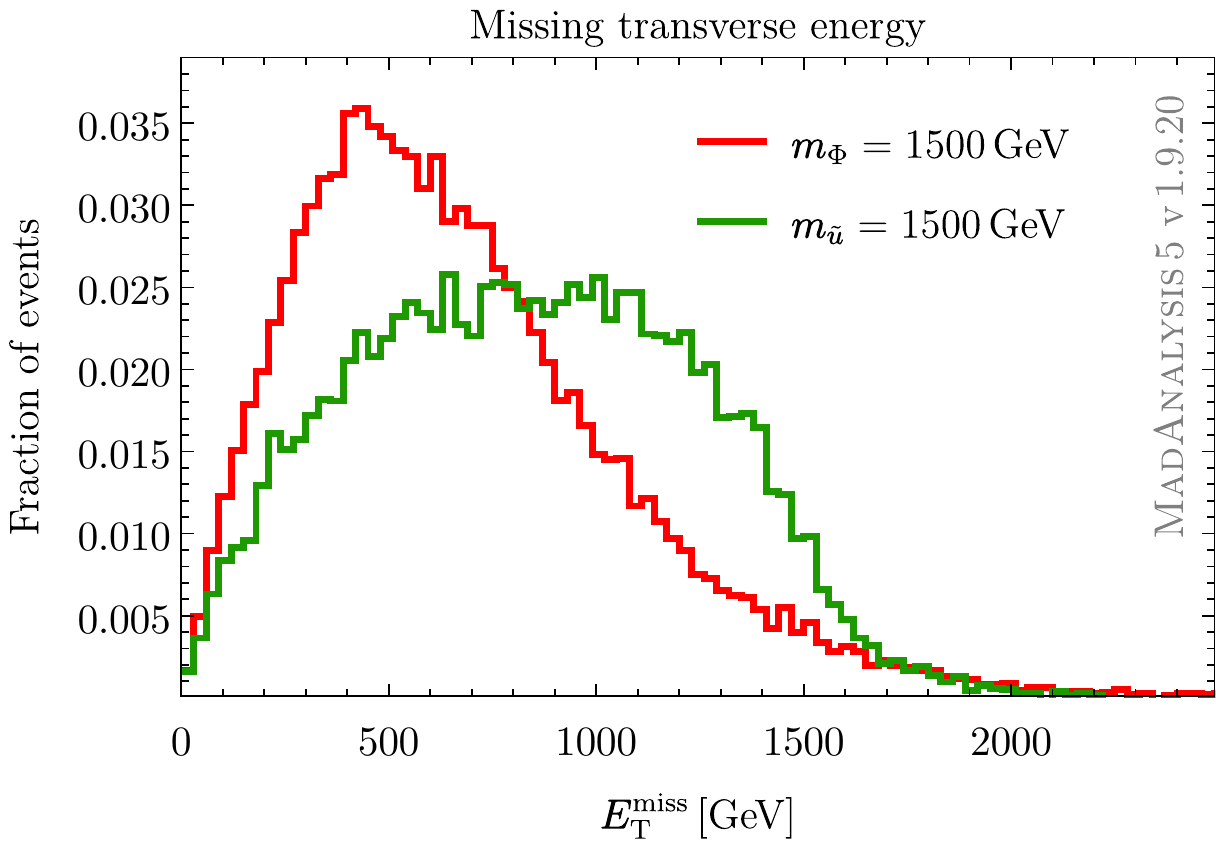}\\
\bigskip
\hspace*{-3.5ex}\includegraphics[scale=0.728]{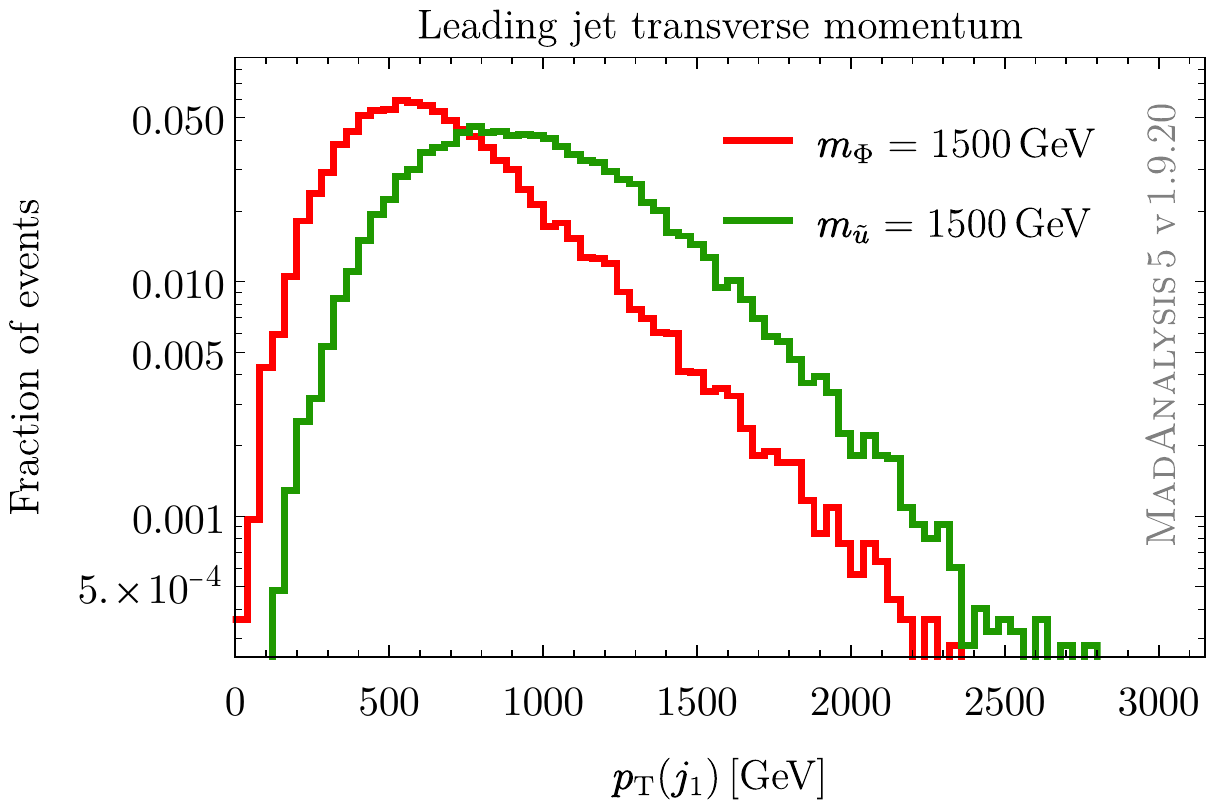}\\
\bigskip
\includegraphics[scale=0.672]{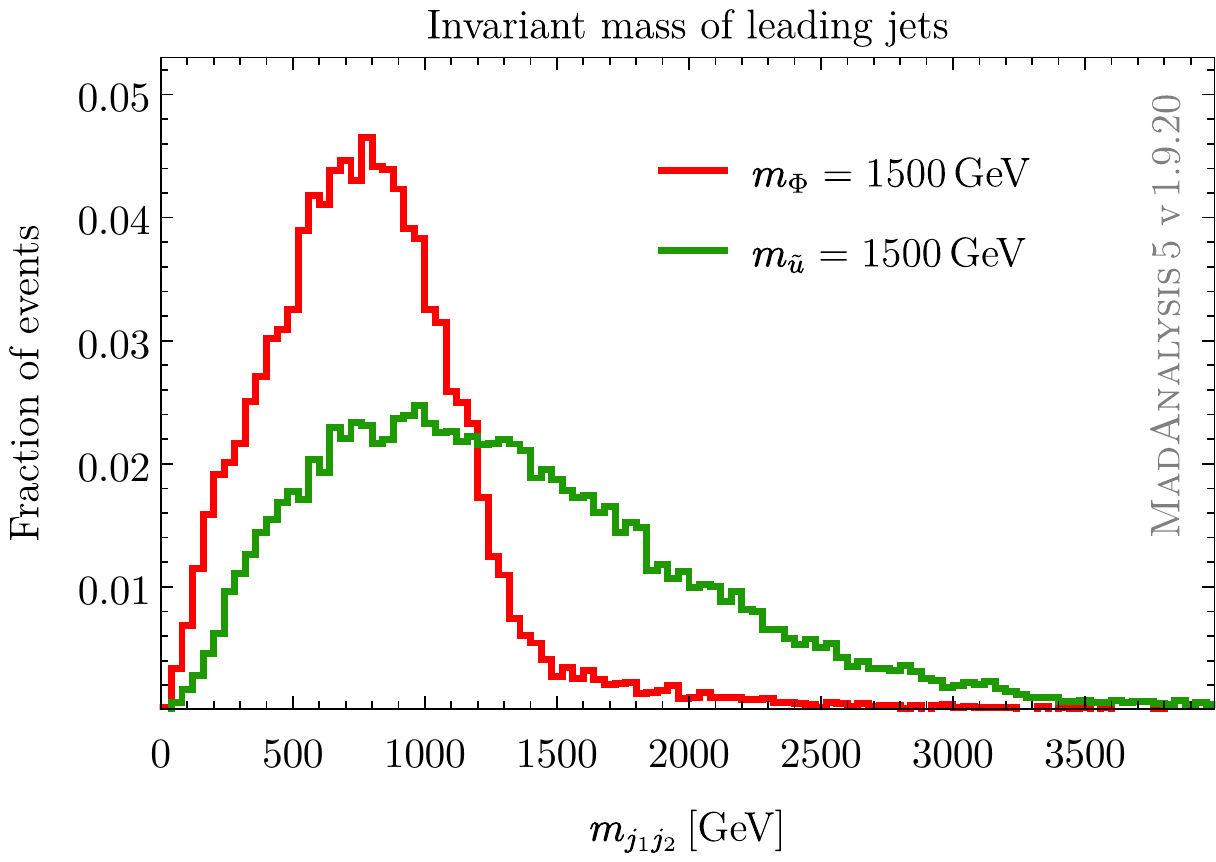}
\caption{Distributions of (top) missing transverse energy, (middle) hardest jet $p_{\text{T}}$, and (bottom) invariant mass of two hardest jets for (red) dimension-six sextet scalar production and (green) first-generation squark pair production at LHC.}
\end{figure}

As for the previous pair of models, we explore the kinematics of these $jj + E_{\text{T}}^{\text{miss}}$ processes in \hyperref[sextetMSSMcompFig]{Figure 8}.
We again only consider couplings between exotic scalars and first-generation SM fermions, so that for instance the MSSM process is left-handed up squark production, $pp \to \tilde{u}_{\text{L}}^{\dagger} \tilde{u}_{\text{L}}$. As mentioned in \hyperref[s3.2]{Section 3.2}, ATLAS and CMS have released comparable limits on light-flavor squarks $\tilde{q}$ based on the full Run 2 dataset, $\mathcal{L} \approx 139\,\text{fb}^{-1}$: ATLAS excludes $m_{\tilde{q}} < 1210\,\text{GeV}$ assuming one non-degenerate light-flavor squark and a light neutralino ($m_{\tilde{q}} < 1850\,\text{GeV}$ for eight degenerate squarks) \cite{ATLAS:2020syg}, and CMS excludes $m_{\tilde{q}} < 1250\,\text{GeV}$ ($1710\,\text{GeV}$) in the same scenarios \cite{CMS:2019ybf}. For the purposes of this discussion, we do not take a firm position on the MSSM squark spectrum and suggest as a starting point some $m_{\tilde{q}}$ lying between the aforementioned limits. $m_{\tilde{u}} = 1500\,\text{GeV}$ happens to reside in this neighborhood, so we use the same scalar masses as in the leptoquark comparison.\footnote{The mass of the neutralino to which $\tilde{u}_{\text{L}}$ decays is set to $m_{\tilde{\chi}^0} = 10\,\text{GeV}$.} The samples were produced and analyzed using the same toolchain as before; the squark sample relied on the MSSM implementation shipped with \textsc{MG5\texttt{\textunderscore}aMC} \cite{MSSM_SLHA2}.

The top panel of \hyperref[sextetMSSMcompFig]{Figure 8} shows the $E_{\text{T}}^{\text{miss}}$ distributions in both models. We see a characteristic peak around $m_{\tilde{u}}/2$ in the MSSM and a broader distribution for the sextet, both because the electron neutrino is lighter than the neutralino in our MSSM benchmark and because one of the neutrinos recoils off of the color-sextet scalar. The middle panel plots the transverse momentum of the leading jet; these distributions are quite similar to the leading-jet $p_{\text{T}}$ distributions in \hyperref[sextetLQcompFig]{Figure 6} since the neutralino is still fairly light. Finally, the bottom panel shows the invariant mass $m_{j_1j_2}$ of the two leading jets, which --- in contrast to the missing energy --- is sharply peaked at $m_{\Phi}/2$ in the sextet model but much broader in the MSSM. Altogether, we again have a pair of scenarios with light exotic particles that must be specified in a renormalizable or effective theory in order to capture the LHC kinematics.
\section{Conclusions}
\label{s5}

We have introduced the Light Exotics Effective field Theory (LEX-EFT) to study the phenomenology of on-shell or nearly on-shell exotic particles. We presented a general method for constructing a complete catalog of operators coupling these new states to the Standard Model. The LEX states are categorized by their SM quantum numbers, and we outlined a general iterative tensor product method to create a complete list of gauge singlets up to the desired mass dimension of effective operator. We described the effect of charge flow on the operator coefficients, which are comprised of distinct products of Clebsch-Gordan coefficients.  We demonstrated through some simple examples that these are important to determining the range of validity of the effective operator, even as they strongly affect production cross sections within the EFT framework.  

We also discussed the distinct kinematics of LEX-EFT operators via an example model of $\mathrm{SU}(3)_{\text{c}}$ color sextets coupling to the Standard Model through dimension-six operators. We showed how several kinematic observables within this model are strongly dependent on the specific LEX state by way of comparison with another model producing in-principle identical final states at LHC. Such distinctive kinematics allow for tailored collider searches more powerful than inclusive searches tuned to models of supersymmetry or leptoquarks. We think this highlights the need for a wider array of collider searches driven by more general models. Finally, we created an example LEX-EFT operator catalog detailing the couplings of a CP-even scalar to pairs of SM gauge bosons up to mass dimension seven. This demonstrated the use of the iterative tensor product method and hinted at the wide array of nonstandard particles that may be accessed through this portal, some of which depart greatly from well trodden bSM paths. We note that in previous work, we presented a catalog of $\mathrm{SU}(2)_{\text{L}}$ singlet color-sextet spin-0 and spin-$\tfrac{1}{2}$ fields up to mass dimension six \cite{Carpenter:2021rkl}. It is remarkable that though both of these endeavors represent only a tiny fraction of the full possible operator catalog, they still yield interesting and surprising interactions between SM and bSM states and spectacular collider phenomenology. 

Opportunities for further work in the LEX-EFT framework are manifold, and we take this opportunity to lay out a long-term plan for the in-depth study of this paradigm. The first and most obvious step is to build out the operator catalog with new LEX states. Approaches to building the operator catalog may follow several systematic paths. One of these, as suggested in this work, is to index this catalog by the SM quantum numbers of the exotic state(s); that is, to specify the representations of the exotic state and use the iterative tensor product procedure to obtain all singlet operators up to the desired mass dimension. Another possible way to aggregate LEX-EFT operators is a portal-based approach. For example, in this work, we specified SM gauge boson pairs as the portal to new physics, and we built a catalog of all possible CP-even scalar LEX fields that can be accessed through this portal through mass dimension seven. These approaches are complimentary: the portal-based exercise gives an idea of which exotic states can be produced through certain processes, and the full study of such states then requires the construction of a complete EFT.

More immediate work might follow directly from topics brought up in this paper. For example, it might be interesting to follow up with collider studies on any of the effective operators in our example catalog in higher-dimensional representations of $\mathrm{SU}(2)_{\text{L}}$, since that promises to yield complex collider signatures. Another route might be to construct the complete operator list up to dimension seven for any of these states, considering all couplings to the SM beyond the diboson portal.

More generally, the kinematic landscape of possible collider final state is vast. We have demonstrated the unique kinematic features that appear in particular models even when final state particles are the same or at least indistinguishable in a detector. In the continuing absence of definitive collider evidence for physics beyond the Standard Model, a systematic search through possible event topologies is needed. Once complete, the LEX-EFT operator catalog can be mined for new collider phenomenology. As demonstrated by the diboson portal presented here and previous work on color sextets, collider final states for the LEX-EFT catalog can have nonstandard collider signatures and striking event topologies that would not be predicted without the catalog. Working through the LEX-EFT catalog scans the landscape of possible collider signatures for new physics, taking a ``leave no stone unturned'' approach.  

Finally, we reiterate that it will eventually be incumbent upon theorists to complete the EFTs for classes of model with particularly compelling phenomenology. This will likely follow the path of recent developments in dark matter studies with simplified models and then next-generation models \cite{LHCDarkMatterWorkingGroup:2018ufk}. This will no doubt offer the benefit of expanding the theoretical landscape of bSM paradigms beyond the standard fare, and may lead to the discovery of new theoretical mechanisms or paradigms.

\acknowledgments

This work was supported by the Department of Physics of The Ohio State University.

\bibliographystyle{Packages/JHEP}
\bibliography{Bibliography/bibliography.bib}

\end{document}